\documentclass[a4paper]{article}
\pdfoutput=1

\usepackage{jheppub}

\usepackage{graphicx}   % need for figures
\usepackage{color}      % use if color is used in text
\usepackage[dvipsnames]{xcolor}      % use if color is used in text
\usepackage{hyperref}   % use for hypertext links, including those to external documents and URLs
\usepackage{slashed}    % for feynmann slashed notation
\usepackage[normalem]{ulem}  % for strike out
\usepackage{textcomp}
\usepackage{tikz}
\usepackage{comment}
\usepackage[utf8]{inputenc} 
\usepackage{bm}
\usepackage{chemfig}
\usepackage{ifthen} % if
\usepackage{soul}

\usetikzlibrary{decorations.shapes}
\usetikzlibrary{decorations.pathreplacing}
\usetikzlibrary{calc}

\newcommand{\jet}{\mathrm{jet}}
\newcommand{\ptj}{p_{T,\jet}}
\newcommand{\mj}{m_{\jet}}

\newcommand\ANN[1]{\mathcal{N}_{#1}}

\newcommand{\HNMR}{${}^1\mathrm{H}$-NMR}

\newcommand{\figref}[1]{Figure \ref{#1}}
\newcommand{\sectionref}[1]{Sec. \ref{#1}}
\renewcommand{\eqref}[1]{Eq. \ref{#1}}

\begin{document}

\title{Spectral Analysis of Jet Substructure with Neural Networks: Boosted Higgs Case}
\author[a]{Sung Hak Lim}
\author[a,b,c]{and Mihoko M. Nojiri}
\affiliation[a]{Theory Center, IPNS, KEK, \\
1-1 Oho, Tsukuba, Ibaraki 305-0801, Japan}
\affiliation[b]{The Graduate University of Advanced Studies (Sokendai), \\
1-1 Oho, Tsukuba, Ibaraki 305-0801, Japan}
\affiliation[c]{Kavli IPMU (WPI), University of Tokyo, \\
5-1-5 Kashiwanoha, Kashiwa, Chiba 277-8583, Japan}


\emailAdd{sunghak.lim@kek.jp, nojiri@post.kek.jp}

\preprint{KEK-TH-2060}

\abstract{
Jets from boosted heavy particles have a typical angular scale which can be used to distinguish them from QCD jets.
We introduce a machine learning strategy for jet substructure analysis using a spectral function on the angular scale. 
The angular spectrum allows us to scan energy deposits over the angle between a pair of particles in a highly visual way.
We set up an artificial neural network (ANN) to find out characteristic shapes of the spectra of the jets from heavy particle decays.
By taking the Higgs jets and QCD jets as examples, we show that the ANN of the angular spectrum input has similar performance to existing taggers. In addition, some improvement is seen when additional extra radiations occur.
Notably, the new algorithm automatically combines the information of the multi-point correlations in the jet.
}

\maketitle
\setcounter{page}{2}
\flushbottom

\section{Introduction}

At multi TeV $pp$ colliders such as the LHC, 
boosted heavy particles can be produced and form a single collimated cluster of particles.
Such a localized cluster is distinguished from a QCD jet from a hard quark or gluon by the substructures of the cluster \cite{Butterworth:2008iy}.
For this purpose, consistent definitions of substructures of jets have been studied extensively.
There are various methods for identifying the jet substructures, such as strategies based on cluster decomposition \cite{Butterworth:2008iy,Thaler:2008ju,Kaplan:2008ie,Plehn:2009rk,Plehn:2010st,Soper:2011cr,Soper:2012pb,Dasgupta:2013ihk,Soper:2014rya,Larkoski:2014wba} and shape variables \cite{Gallicchio:2010sw,Thaler:2010tr,Gallicchio:2011xq,Chien:2013kca,Larkoski:2013eya,Larkoski:2014gra,Moult:2016cvt}. 
These methods focus on different features of jet substructures to maximize the discrimination power.
For the case of Higgs, W, and Z boson decaying hadronically into two quarks, a critical feature is a two-prong substructure inside.
Because the key features depend on the nature of the parent particle of a jet, there are several frameworks that can be applied to jets \cite{Jankowiak:2011qa,Datta:2017rhs,Chakraborty:2017mbz,Aguilar-Saavedra:2017rzt,Komiske:2017aww,Chien:2018dfn}.

We propose a spectral analysis in order to identify originating partons of a jet using a spectral function inspired by the proton nuclear magnetic (\HNMR) spectroscopy of organic molecules. 
The organic molecules consist mostly of a carbon skeleton and hydrogen atoms.
These substructures can be identified by the \HNMR~spectrum, which records the resonant frequency of the hydrogen nuclei under an external magnetic field.
The resonant frequency depends on an induced magnetic field generated by the rest of the molecular substructure; hence, the molecular structures can be determined by investigating the spectrum, as shown in Figure 1.
In particular, we use shift and splitting of the resonance frequency, where the big shift comes from the electron cloud, whose density is determined by the skeletal structure, and small splittings are from the electromagnetic interaction between the hydrogen nuclei.
Similarly, we develop a spectral function for jet substructure study, where the big structure of the spectrum is made from initiating hard partons while small structures come from QCD interaction from the hard partons.
The spectral function that we propose is similar to the angular structure function \cite{Jankowiak:2011qa,Jankowiak:2012na,Larkoski:2012eh}. 
The resulting spectrum contains useful information for identifying the nature of a given jet.

\begin{figure}
\begin{center}
\begin{tabular}{rcc}
\chemfig{
{C}
(-[:210]\textcolor{red}{H})
(<[:70]\textcolor{red}{H})
(<:[:110]\textcolor{red}{H})
(-[:330]{C}(
	(<[:290]\textcolor{blue}{H})
	(<:[:250]\textcolor{blue}{H})
)-[:30]O\textcolor{OliveGreen}{H})
}
&
\begin{tikzpicture}[baseline={([yshift=-.5ex]current bounding box.center)},vertex/.style={anchor=base,
    circle,fill=black!25,minimum size=18pt,inner sep=2pt},scale=1.0]
\draw [->] (0,0.1) -- (0.7,0.1);
\draw [<-] (0,-0.1) -- (0.7,-0.1);
\end{tikzpicture}
&
\begin{tikzpicture}[baseline={([yshift=0.ex]current bounding box.center)},vertex/.style={anchor=base,
    circle,fill=black!25,minimum size=18pt,inner sep=2pt},xscale=0.90,yscale=0.65]
\draw (0.5,0) -- (-4.5,0);
% OH
\node [draw=none,anchor=south] at (-2.61,1.3) {\sout{\phantom{0}}$\mathrm{O{\color{OliveGreen}{H}}}$};
\draw [ultra thick, color=OliveGreen](-2.61, 0) -- (-2.61,1.3);
% -CH2- 3.687
\node [draw=none,anchor=south] at (-3.687,2.0) {\sout{\phantom{0}}$\mathrm{C{\color{blue}{H_2}}}$\sout{\phantom{0}}};
\draw [color=blue] (-3.537, 0) -- (-3.537, 1);
\draw [color=blue] (-3.637, 0) -- (-3.637, 2);
\draw [color=blue] (-3.737, 0) -- (-3.737, 2);
\draw [color=blue] (-3.837, 0) -- (-3.837, 1);
% -CH3 1.226
\node [draw=none,anchor=west] at (-1.126,2.0) {\sout{\phantom{0}}$\mathrm{C{\color{red}{H_3}}}$};
\draw [color=red] (-1.126, 0) -- (-1.126, 2.25);
\draw [color=red] (-1.226, 0) -- (-1.226, 4.5);
\draw [color=red] (-1.326, 0) -- (-1.326, 2.25);
% ticks
\draw (0,0) -- (0,-0.2);
\draw (-0.5,0) -- (-0.5,-0.1);
\node [draw=none, anchor=north] at (0,-0.2) {0};
\draw (-1,0) -- (-1,-0.2);
\draw (-1.5,0) -- (-1.5,-0.1);
\node [draw=none, anchor=north] at (-1,-0.2) {1};
\draw (-2,0) -- (-2,-0.2);
\draw (-2.5,0) -- (-2.5,-0.1);
\node [draw=none, anchor=north] at (-2,-0.2) {2};
\draw (-3,0) -- (-3,-0.2);
\draw (-3.5,0) -- (-3.5,-0.1);
\node [draw=none, anchor=north] at (-3,-0.2) {3};
\draw (-4,0) -- (-4,-0.2);
\draw (-4.5,0) -- (-4.5,-0.1);
\node [draw=none, anchor=north] at (-4,-0.2) {4};
\node [draw=none, anchor=north] at (-2,-1.0) {chemical shift};
\end{tikzpicture}
\end{tabular}
\end{center}
\caption{
Molecular structure of ethanol and its \HNMR~spectrum. The intensity, location and splitting of peaks allow us to identify the original molecular structure. The chemical shift is the resonant frequency of a hydrogen nucleus relative to a reference frequency.
}
\label{fig:spectrum_analogy_from_OCHEM}
\end{figure}
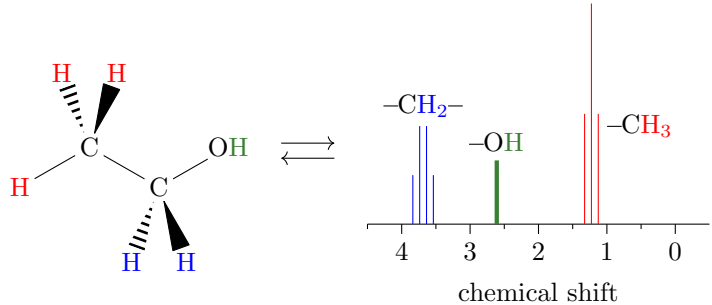

An artificial neural network (ANN) is a useful model for analyzing the spectrum.
Jet substructure analyses based on ANN are gaining attention recently and have been studied in various contexts.
The analyses are categorized mainly into two groups with different inputs.
One group utilizes special-purpose observables and uses ANN to identify a correlation between substructures \cite{Datta:2017rhs,Aguilar-Saavedra:2017rzt,Datta:2017lxt} similar to analyses with a boosted decision tree \cite{Bhattacherjee:2016bpy,Chien:2017xrb}.
The other group uses the jet constituents directly and uses ANN to find out particular substructures in a jet. This group is again categorized into two subgroups depending on how you interpret jet constituents.
One can interpret jet constituents as an image \cite{Cogan:2014oua} and use image recognition techniques \cite{Almeida:2015jua, deOliveira:2015xxd,Baldi:2016fql,Komiske:2016rsd,Kasieczka:2017nvn,Choi:2018dag}. 
The other strategies are to interpret a jet as a sequence of data, such as clustering sequence of the jet algorithms \cite{Bartel:1986ua,Catani:1993hr,Ellis:1993tq,Dokshitzer:1997in,Wobisch:1998wt,Cacciari:2008gp}, and utilize ANN for sequential data anlysis \cite{Louppe:2017ipp,Cheng:2017rdo,Egan:2017ojy}
Our approach is different from these approaches, namely our network is requested to analyze an event-by-event spectrum of a jet.
This approach reduces the inputs of the ANN significantly but it can still learn characteristic non-local correlations in a jet from a heavy particle like the non-local neural networks for video classification \cite{NonLocal2018}.
We will show that our approach improves the separation between Higgs jets and QCD jets in natural manner.

This paper is organized as follows. 
In \sectionref{sec:2}, we define a spectral function $S_2(R)$ and describe its nature.
We also explain the setup of our Monte-Carlo simulations.
In \sectionref{sec:3}, we show a cut-based classification of the Higgs jet vs QCD jet using the spectrum $S_2(R)$ and compare the performance with the ratio of the enegy correlation function, $D_2$ \cite{Larkoski:2014gra}, for two-prong substructures.
In \sectionref{sec:4}, we introduce the spectral analysis of jet substructure with ANN.
\sectionref{sec:5} is devoted for summary and discussions.

\section{A Spectral Function for Jet Substructure}
\label{sec:2}

The jet substructure analysis has similarities with the \HNMR~analysis of organic molecular structures. 
An organic molecule has a carbon skeleton surrounded by hydrogens, which \HNMR~analyzes to find out the topological skeleton.
On the other hand, a jet of our interest arises from hard partons, which can be regarded as a topological skeleton of the momentum distribution of jet constituents.
The jet constituents radiated from hard partons are something analogous to the hydrogens in a organic molecule.

In the \HNMR~analysis, we record interactions between hydrogen nuclei and the rest of the molecular structure by their resonant frequencies to measure the molecular structure.
Likewise, we focus on correlations between pairs of the constituents of the jet based on distance to identify the originating partons.
A popular choice of the distance is angular distance $R_{ij} = \sqrt{\eta^2_{ij} + \phi^2_{ij}}$, where $\eta_{ij}$ is pseudorapidity and $\phi_{ij}$ is azimuthal angle between the jet constituents $i$ and $j$.
Hence, we define a binned spectral function of angular distance $R$ as follows,
\begin{eqnarray}
S_2 (R; \Delta R) & = & \frac{1}{\Delta R} \sum_{ \substack{ i,j \in \jet \\ R_{ij} \in [R ,R + \Delta R)  }} p_{T,i} p_{T,j} ,
\label{eqn:def_s2}
\end{eqnarray}
where $\Delta R$ is a bin width, and $p_{T,i}$ and $p_{T,j}$ are transverse momenta of jet constituents $i$ and $j$.
In a continuum limit $\Delta R \rightarrow 0$,\footnote{For more formal description, see \cite{Tkachov:1995kk}.} this binned spectral function turns into, 
\begin{eqnarray}
S_2 (R) 
& = &
\int d \vec{R}_1 d \vec{R}_2 \; P_T(\vec{R}_1) P_T(\vec{R}_2) \cdot \delta( R - R_{12} ),
\\
P_T(\vec{R})
& = &
\sum_{\substack{i \in \jet}} p_{T,i} \delta( \vec{R} - \vec{R}_i )
\label{eqn:def_s2_cont}
\end{eqnarray}
where $d \vec{R} \, P_T(\vec{R})$ is a $p_T$ sum of constituents in a neighborhood $d \vec{R}$ of $\vec{R}$, and $\delta(x)$ is the Dirac $\delta$ function.
Integrating $S_2(R)$ over the bin range $[R,R+\Delta R)$ returns the binned spectral function, 
\begin{equation}
S_2 (R; \Delta R) = \frac{1}{\Delta R} \int_R^{R+\Delta R} d R \, S_2 (R).
\end{equation}
The two-point correlation spectral function may be easily generalized to three-point or multi-point correlation spectral function, analogous to the energy correlation functions \cite{Larkoski:2013eya} and the energy flow polynomials \cite{Komiske:2017aww}.
However, those generalizations are out of the scope of this paper.

The spectral function is infrared and collinear (IRC) safe, namely invariant under soft and collinear radiations.
If the IRC safety is not satisfied, Kinoshita-Lee-Nauenberg theorem \cite{Kinoshita:1962ur,Lee:1964is} is not applicable, and the resulting spectrum is hard to be estimated from perturbative QCD calculations. 
Soft radiation does not change $S_2(R)$ because the soft radiation has a zero transverse momentum, which has no impact on $P_T(\vec{R})$ as well as $S_2(R)$.
Collinear radiation does not change $S_2(R)$ because the products stay at the original $\vec{R}$ coordinate.
The momenta of the products are added together at $\vec{R}$; therefore, $P_T(\vec{R})$ and $S_2(R)$ are invariant.

The IRC safety of the spectral function can be understood easily by explicit examples such as a jet with a single constituent.
Suppose that $a$ is the only jet constituent.
This jet has the only angular scale $R_{aa}=0$ and its binned spectrum is given as follows,
\begin{eqnarray}
\label{eqn:s2_jet_single}
S_2(R; \Delta R)
& = &
\frac{1}{\Delta R} 
\begin{cases}
p_{T,a}^2   & \mathrm{if}\;R = 0, \\
0         & \mathrm{if}\;R \neq 0.
\end{cases}
\end{eqnarray}
Invariance of this $S_2(R; \Delta R)$ under soft radiation is trivial, and hence, we consider collinear splitting of $a$.
Suppose that $b$ and $c$ are the jet constituents from the collinear splitting of $a$.
Then all the pairs $bb$, $bc$, $cb$, and $cc$ have the same angular scale $R_{ij}=0$. 
If $b$ and $c$ carry $z$ and $1-z$ fraction of the transverse momenta $p_{T,a}$, $S_2(R;\Delta R)$ turns into
\begin{eqnarray}
\label{eqn:s2_jet_single_coll_split}
S_2(R; \Delta R)
& = &
\frac{1}{\Delta R}
\begin{cases}
z^2 p_{T,a}^2 + z(1-z) p_{T,a}^2 + (1-z) z p_{T,a}^2 + (1-z)^2 p_{T,a}^2  & \mathrm{if}\;R = 0, \\
0         & \mathrm{if}\;R \neq 0,
\end{cases}
\\
& = &
\frac{1}{\Delta R} 
\begin{cases}
p_{T,a}^2   & \mathrm{if}\;R = 0, \\
0         & \mathrm{if}\;R \neq 0.
\end{cases}
\end{eqnarray}
Therefore, the binned spectrum is IRC safe.
Note that summing the autocorrelation term $p_{T,i}^2$ in \eqref{eqn:def_s2} is necessary to achieve IRC safety at $R=0$, because the crossing term $z(1-z) p_{T,a}^2$ after the splitting is originated from the autocorrelation term $p_{T,a}^2$.

We show another example of a jet with two constituents $a$ and $b$, so that it has a non-zero spectrum at $R = R_{ab} > 0 $.
Now the binned spectrum has cross-correlation terms at non-zero angular scale,
\begin{eqnarray}
S_2(R; \Delta R)
& = &
\frac{1}{\Delta R} 
\begin{cases}
p_{T,a}^2 + p_{T,b}^2   & \mathrm{if}\;R = 0, \\
0                       & \mathrm{otherwise},
\end{cases}
+
\frac{1}{\Delta R}
\begin{cases}
2 p_{T,a} p_{T,b}   & \mathrm{if}\;R_{ab} \in [R, R+\Delta R), \\
0                       & \mathrm{otherwise}.
\end{cases}
\label{eqn:s2_jet_double}
\end{eqnarray}
Suppose that a collinear splitting of $b$ produces two partons with transverse momenta $z p_{T,b}$ and $(1-z) p_{T,b}$ respectively.
Then the binned spectrum turns into
\begin{eqnarray}
\nonumber
S_2(R; \Delta R)
& = &
\frac{1}{\Delta R} 
\begin{cases}
p_{T,a}^2 + \left[ z^2 p_{T,b}^2 + z(1-z) p_{T,b}^2 + (1-z) z p_{T,b}^2 + (1-z)^2 p_{T,b}^2 \right]    & \mathrm{if}\;R = 0, \\
0                       & \mathrm{otherwise,}
\end{cases}
\\ & &
+
\frac{1}{\Delta R}
\begin{cases}
2 z p_{T,a} p_{T,b} + 2 (1-z) p_{T,a} p_{T,b}   & \mathrm{if}\;R_{ab} \in [R, R+\Delta R), \\
0                                               & \mathrm{otherwise},
\end{cases}
\\
& = &
\frac{1}{\Delta R} 
\begin{cases}
p_{T,a}^2 + p_{T,b}^2   & \mathrm{if}\;R = 0, \\
0                       & \mathrm{otherwise},
\end{cases}
+
\frac{1}{\Delta R}
\begin{cases}
2 p_{T,a} p_{T,b}   & \mathrm{if}\;R_{ab} \in [R, R+\Delta R), \\
0                       & \mathrm{otherwise}.
\end{cases}
\end{eqnarray}
Therefore, the binned spectrum is IRC safe.
In general, this IRC safety is achieved by the bilinear term of $p_{T,i}$ and $p_{T,j}$ in \eqref{eqn:def_s2} like the other jet substructure variables directly built from jet constituents \cite{Gallicchio:2010sw,Thaler:2010tr,Gallicchio:2011xq,Jankowiak:2011qa,Larkoski:2013eya,Larkoski:2014gra,Moult:2016cvt,Komiske:2017aww}.

The IRC safety of the spectral function $S_2(R)$ is also understood in the context of $C$-correlators \cite{Tkachov:1995kk,Komiske:2017aww}.
The $S_2(R)$ is a special case of $C$-correlators with an unbounded non-smooth angular weighting function $f_2 (\hat{p}_1, \hat{p}_2) = \delta( R - R_{12} )$.
If we replace the Dirac $\delta$ function to a bounded smoooth function, for example, $\delta (x) \rightarrow (|a| \sqrt{\pi})^{-1} e^{-(x/a)^2}$, the Taylor expansion of $\delta (x)$ transforms $S_2(R)$ into a series of IRC safe energy flow polynomials \cite{Komiske:2017aww} with two vertices. 
The series converges to $S_2(R)$ in the limit $a \rightarrow 0$, and the IRC safety of the spectral function is understood asymptotically. 

Note that the spectral function $S_2(R)$ is a basis of bilinear $C$-correlators $F_2$ \cite{Tkachov:1995kk} with an angular weighting function $f_2(R_{ij})$ of the angular distance $R_{ij}$,
\begin{equation}
F_2
=
\int d \vec{R}_1 d \vec{R}_2 \; P_T(\vec{R}_1) P_T(\vec{R}_2) \cdot f_2( R_{12} )
=
\int_0^\infty  d R \; S_2(R) \,f_2(R) .
\end{equation}
For example, the zeroth and the second moment of $S_2(R)$ are the one-point and the two-point energy correlation functions \cite{Larkoski:2013eya}, which are approximately the transverse momentum $\ptj$ and the mass $\mj$ of the jet respectively,
\begin{eqnarray}
\label{eqn:S2_sumrule_pt}
\int_0^\infty  dR\, S_2(R) \, \hphantom{R^2} 
& = & 
\makebox[0pt][l]{$\displaystyle \left( \sum_{i\in\jet} p_{T,i} \right)^2 $} 
\phantom{\sum_{i,j\in\jet} p_{T,i} p_{T,j} R_{ij}^2}
\approx \ptj^2 ,\\
\label{eqn:S2_sumrule_mass}
\int_0^\infty  dR\, S_2(R) \, R^2 & =  &\sum_{i,j\in\jet} p_{T,i} p_{T,j} R_{ij}^2
\approx 2 \mj^2 .
\end{eqnarray}
These integrals help to interpret $S_2(R)$.
The spectral densities $S_2(R)$ and $S_2(R)\,R^2 $ measure contribution to $\ptj^2$ and $\mj^2$ from the pairs of jet constituents at the angular scale $R$, respectively.

We perform a Monte Carlo study of Higgs jets vs QCD jets classification using the  spectrum $S_2(R)$.
We generate $pp\rightarrow Zj$ events
and $pp\rightarrow Z h$ events followed by $h\rightarrow b\bar{b}$, and use the leading jet of the events as training sample of one prong and two prong jets respectively.
Each sample is generated at the leading order in QCD using \texttt{MadGraph5\_aMC@NLO 2.6.1} \cite{Alwall:2014hca} with parton distribution function (PDF) set NNPDF 2.3 LO at $\alpha_S(m_Z) = 0.130$ \cite{Ball:2012cx}.
$Z$ bosons are forced to decay into neutrinos so that they are not detected.
The events are showered and hadronized by \texttt{Pythia 8.226} \cite{Sjostrand:2014zea} with Monash tune \cite{Skands:2014pea}. 
We include effects of underlying events such as multi-parton interaction and beam remnant treatment but we do not take pile-ups into account.

Finally, we simulate detector response using \texttt{Delphes 3.3.3} \cite{deFavereau:2013fsa} with their default ATLAS configuration.
Jets are reconstucted from calorimeter towers using anti-$k_T$ algorithm \cite{Cacciari:2008gp} with a jet radius parameter $R_{\jet} =  1.0$ implemented in fastjet 3.3.0 \cite{Cacciari:2011ma,Cacciari:2005hq}.
We study substructures of the leading jets with $\ptj \in [300, 400]$ GeV and $\mj \in [100, 150]$.
The characteristic angle between two $b$ quarks from the boosted Higgs boson is then $R_{b\bar{b}} \gtrsim 2m_h / p_{T,\mathrm{jet}} \approx 0.83 $. 
Hence, the choice of the jet radius is enough to catch parton showers from those two $b$ quarks efficiently.
For $Zh$ events, we additionally require that at least there is one $b$ parton whose momentum in matrix element level is located within $R \leq 1$ from the jet center to remove events with hard initial state radiations.
After these preselections, we have 256691 $Zh$ and $Zj$ events for training ANN.
When we test ANN, we use a testing sample generated independently to the training sample.

\begin{figure}
\begin{center}
\includegraphics[width=0.3\textwidth]{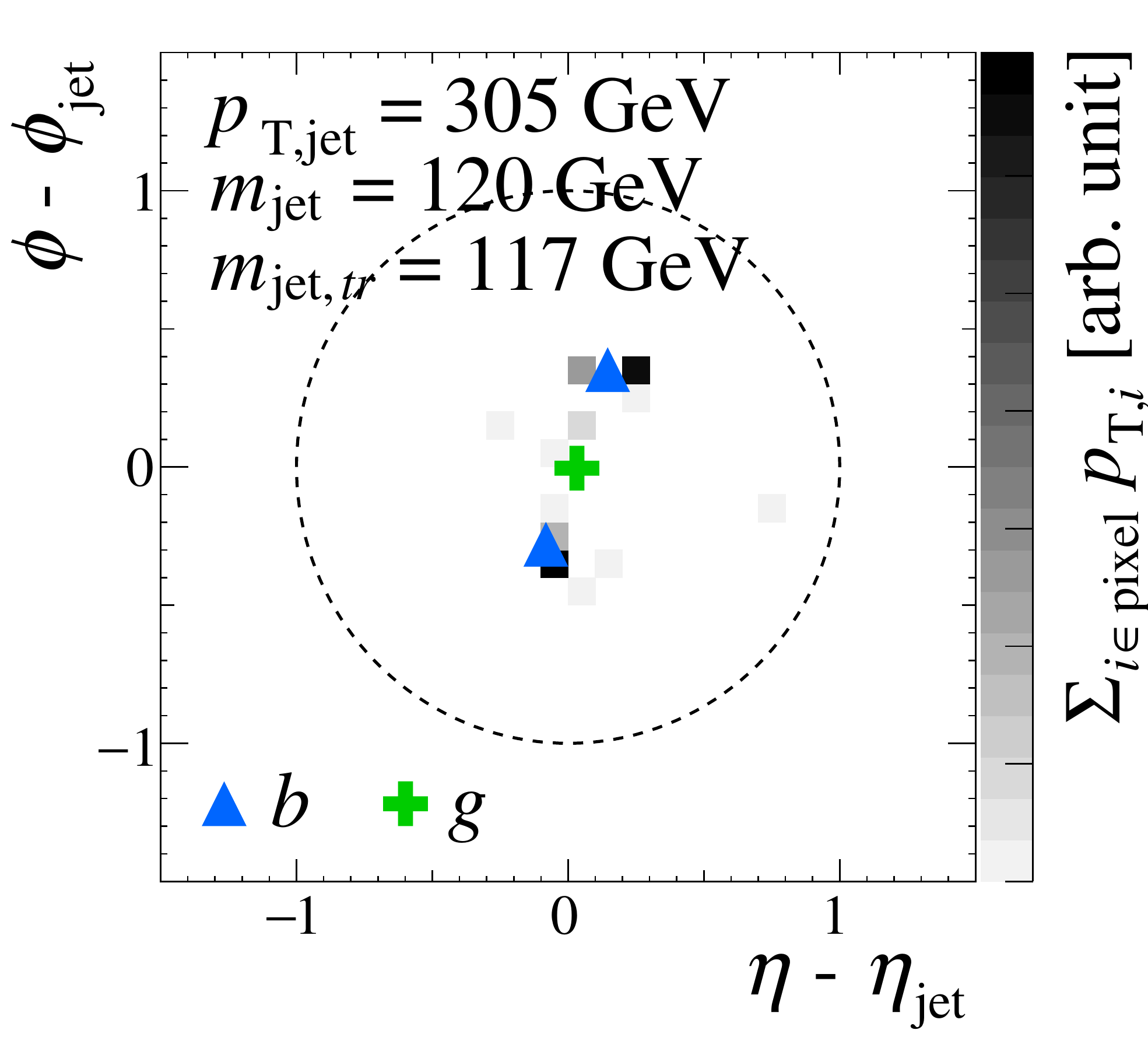}	
\includegraphics[width=0.3\textwidth]{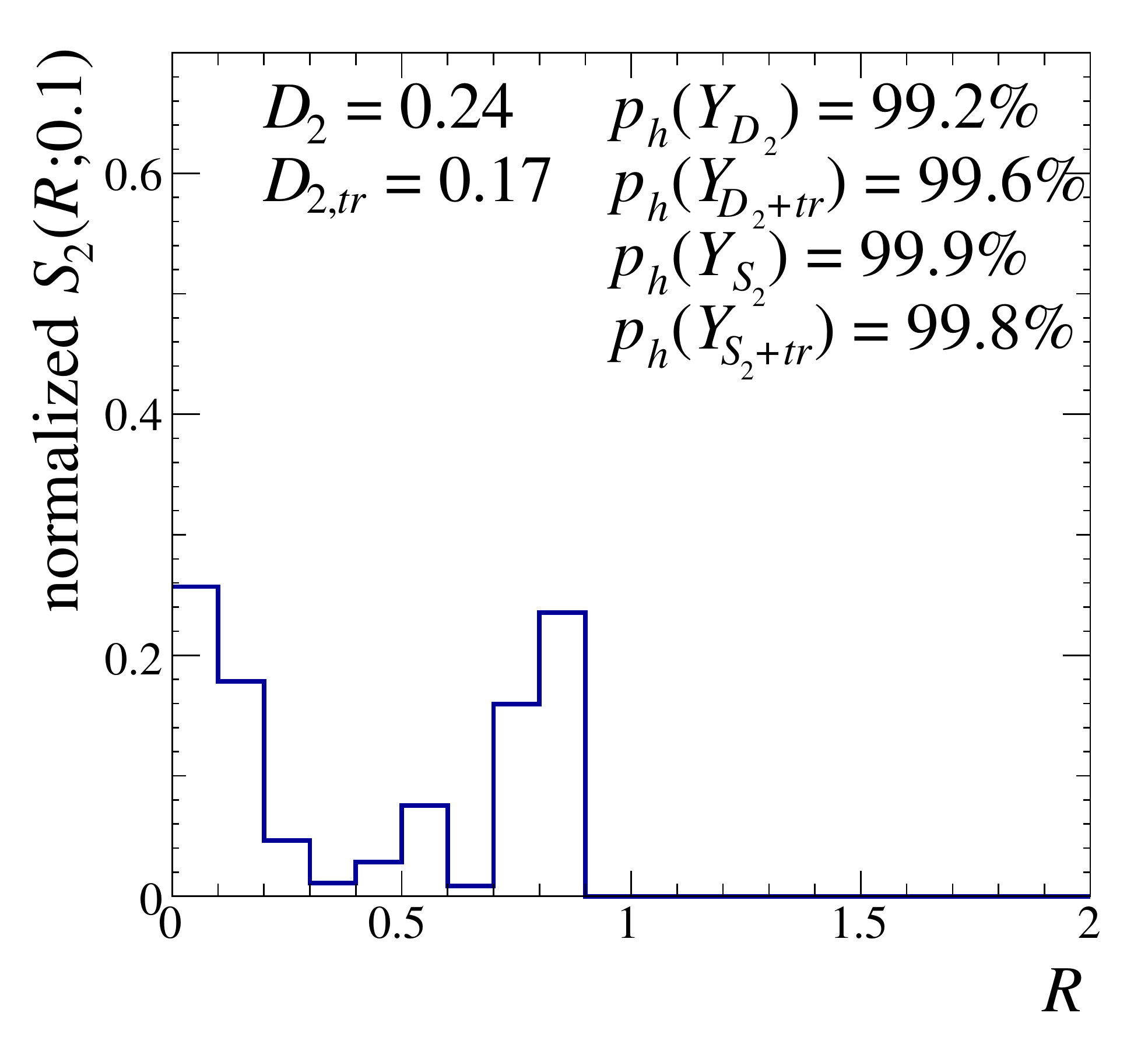}	
\includegraphics[width=0.3\textwidth]{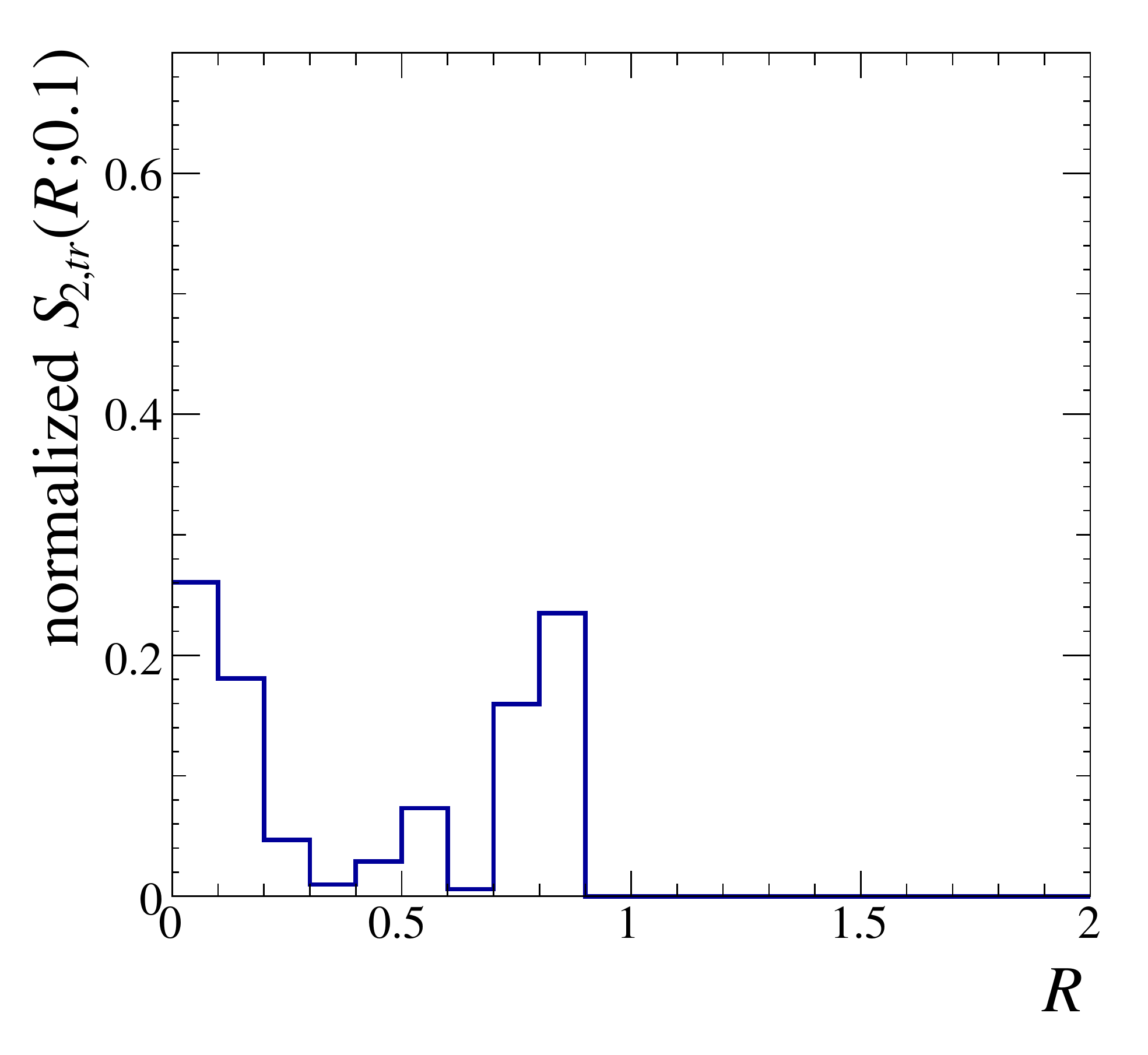}	
\end{center}
\begin{center}
\includegraphics[width=0.3\textwidth]{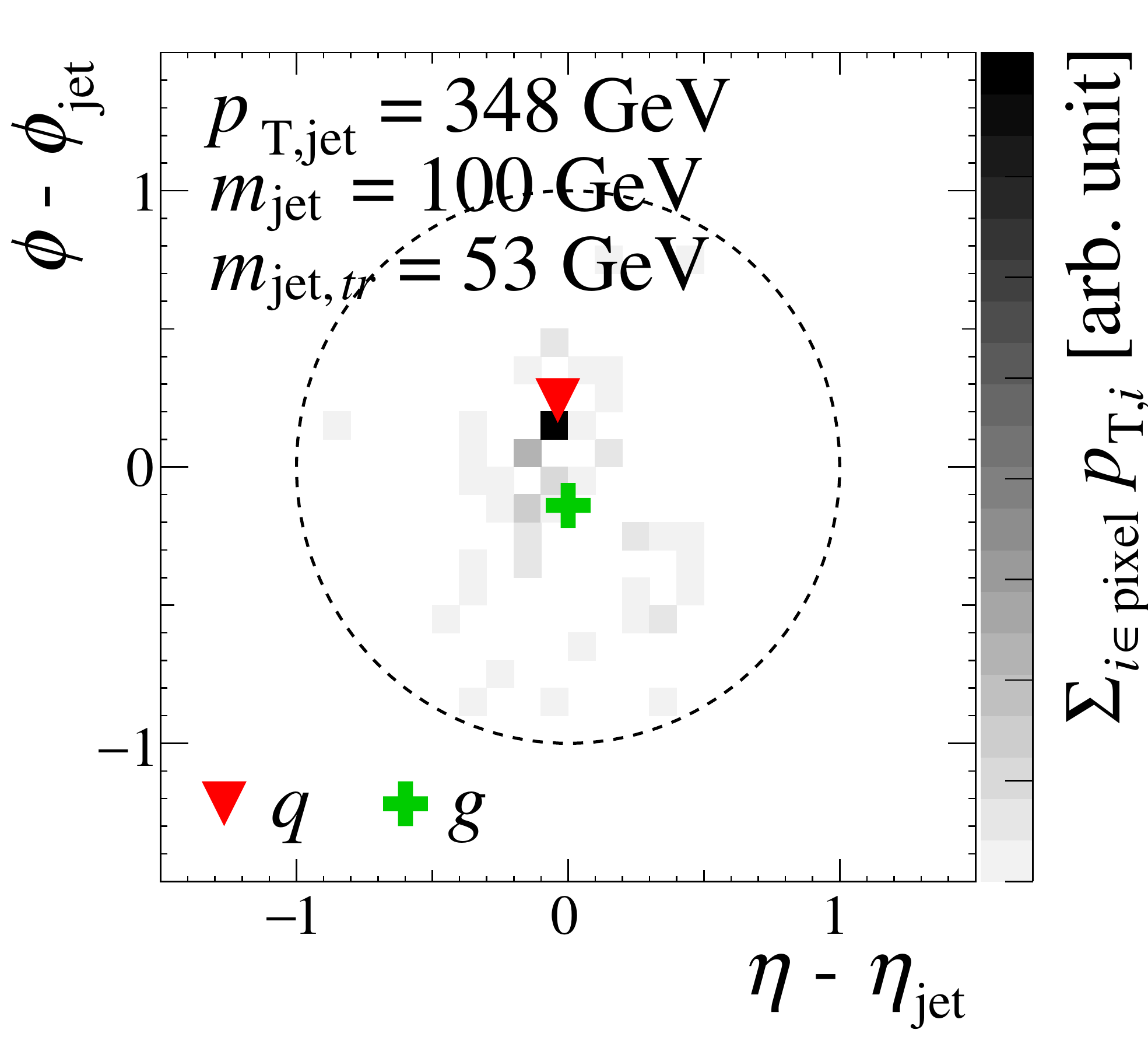}	
\includegraphics[width=0.3\textwidth]{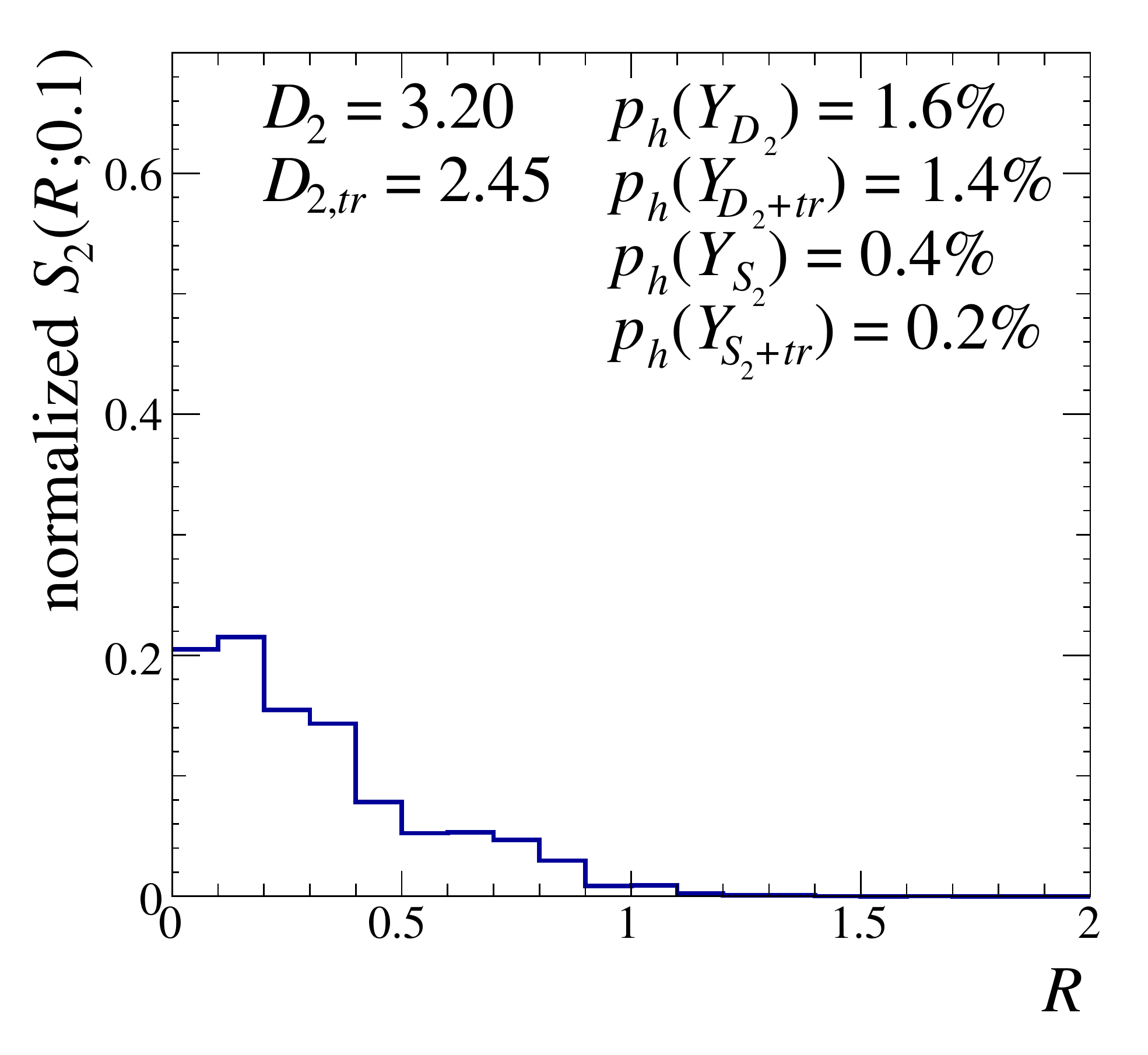}	
\includegraphics[width=0.3\textwidth]{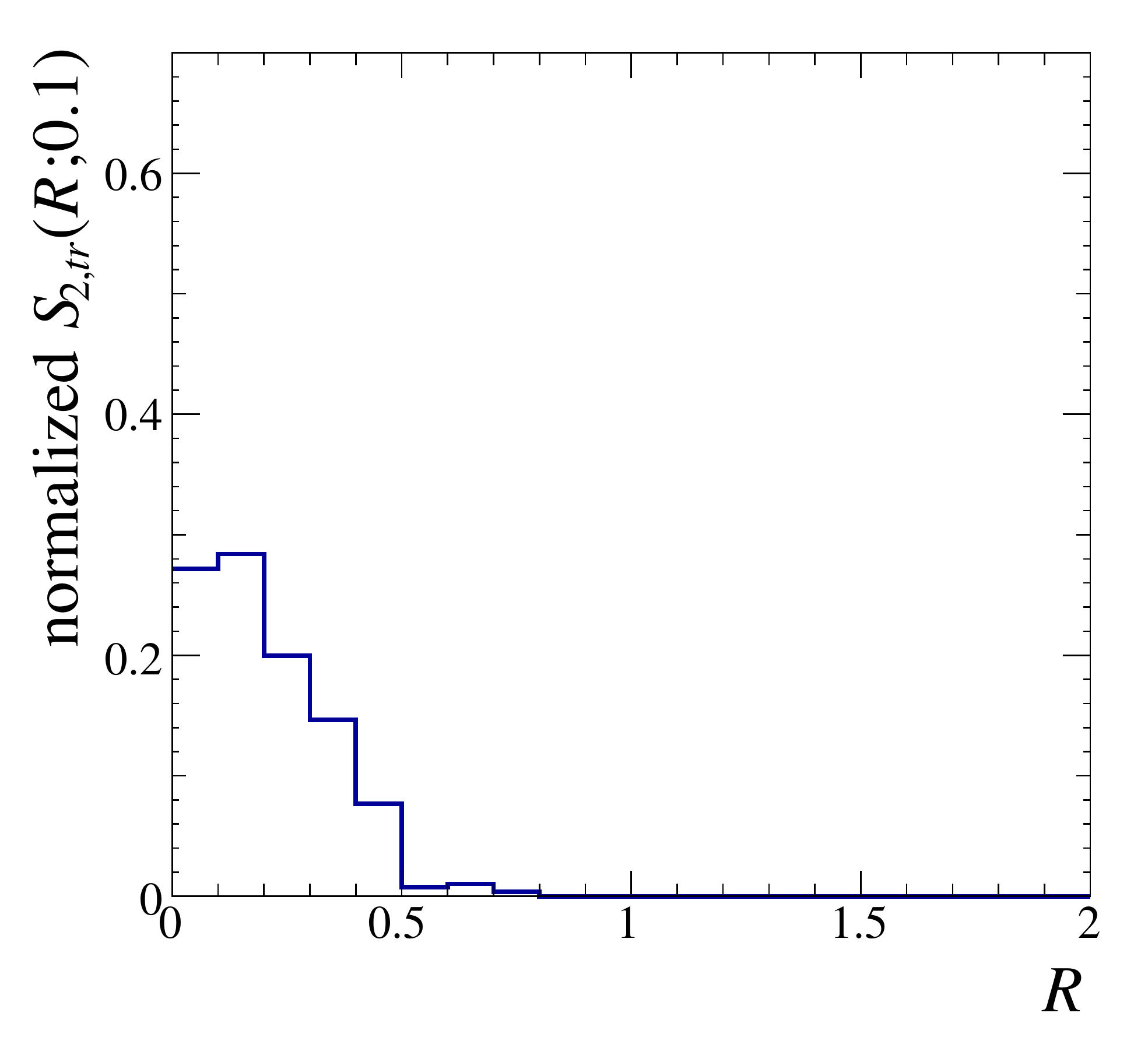}	
\end{center}
\caption{
The pixelated jet image (left), two-point spectrum $S_2(R;0.1)$ (center), and the trimmed spectrum $S_2(R;0.1)$ (right) of a typical Higgs jet (top) and a typical quark jet (bottom).
The sum of the bins of each  spectrum is normalized to 1. 
The red reversed triangles, blue triangles, and green plus symbols in the pixelated jet images indicate direction of light quarks, $b$ quarks and gluons obtained from the matrix element including order $\alpha_S$ radiations.
We show the Higgs-like probabilities of the jets, $p_h(Y_X)$ of an ANN model with various input $X$, defined in \sectionref{sec:4}, in the central plots.
}
\label{fig:jet_profile_char}
\end{figure}

We show typical pixelated jet images and spectra $S_2(R;0.1)$ of a Higgs boson jet and a quark jet in \figref{fig:jet_profile_char}. 
For the boosted Higgs boson, the $S_2(R;0.1)$ distribution has two prominent peaks at $R = 0$ and $R_{b\bar{b}}$ which correspond to autocorrelation and cross-correlation of two $b$ partons respectively,
\begin{eqnarray}
\label{eqn:spectrum2_sig}
p_T(\vec{R})
& = &
p_{T,b} \cdot \delta(\vec{R} - \vec{R}_{b} ) 
+ p_{T,\bar{b}} \cdot \delta(\vec{R} - \vec{R}_{\bar{b}} ) ,
\\ \nonumber
S_2(R)
& = &
\left( p_{T,b}^2 + p_{T,\bar{b}}^2 \right) \cdot \delta(R) + 2 p_{T,b} p_{T,\bar{b}} \cdot \delta( R - R_{b\bar{b}}).
\end{eqnarray} 
Because the Higgs boson decays spherically, the two peaks tend to have comparable intensities, $p_{T,b}^2 + p_{T,\bar{b}}^2 \simeq 2 p_{T,b} p_{T,\bar{b}}$, as shown in \figref{fig:jet_profile_char}.
Parton shower develops along the partons.
Each splitting of a parton is characterized by the angle betwen daughter partons and their momenta.
The spectral density $S_2(R)$ sums up those individual splittings. 
The peaks in $R$ is smeared by the parton shower and hadronization, but it does not change the initial radiation pattern.
In \figref{fig:jet_profile_char}, we also show the quark jet spectrum.
The spectrum does not have distinctive peaks significantly.
Instead, it is gradually decreasing as $R$ increases.

Note that the spectral function $S_2(R)$ is not completely independent to the angular structure function $\Delta \mathcal{G}(R)$ in \cite{Jankowiak:2011qa}.
We have a relation between $S_2(R)$ and $\Delta \mathcal{G}(R)$ as follows,
\begin{eqnarray}
\mathcal{G}(R) 
& = & \frac{ \int_0^R dR' \, S_2(R') \, R'^2  } { \int_0^\infty dR' \,  S_2(R') \, R'^2 },
\\
\Delta \mathcal{G}(R)  
& = & \frac{d \log \mathcal{G}(R)}{d \log R} 
=
\frac{R \cdot S_2(R)\, R^2 }{ \int_0^R dR' \, S_2(R') \,  R'^2} . 
\end{eqnarray}
This $\Delta \mathcal{G}(R)$ is a Higuchi's fractal dimension \cite{HIGUCHI1988277} of $\mathcal{G}(R)$ which measures irregularity of $\mathcal{G}(R)$ over $R$. 
QCD jets have a uniform $\Delta \mathcal{G}(R)$ distribution on average because of approximate scale invariance of QCD \cite{Jankowiak:2011qa,Jankowiak:2012na,Larkoski:2012eh}.
On the other hand, $\mathcal{G}(R)$ of the multi-prong jet shows sharp peaks at some angular scales \cite{Jankowiak:2011qa}.
Hence, a number of peaks and peak heights in $\Delta \mathcal{G}(R)$ can be used as a classifier of the jets.

In the following sections, we will show two Higgs jet classifiers using the spectrum $S_2(R)$.
In \sectionref{sec:3}, we show a cut-based analysis using $S_2(R)$.
In \sectionref{sec:4}, we introduce neural network classifiers using $S_2(R)$.

\section{Cut-Based Spectral Analysis}
\label{sec:3}

Before presenting our neural network classifier for the Higgs jets, we show a cut-based analysis using $S_2(R)$ to get a quick insight on the spectrum. 
Namely, we introduce a ratio of the activity on the characteristic angular scale $\hat{R}_{b\bar{b}}$ of a Higgs jet and that of the surrounding angular scales,
\begin{equation}
R_{S_2} = \frac{\int_{a \hat{R}_{b\bar{b}}}^{\min [ a' \hat{R}_{b\bar{b}}, R_{\jet} ]} d R \, S_2(R)}{ \int_0^{a \hat{R}_{b\bar{b}}} d R \, S_2(R) + \int_{\min [ a' \hat{R}_{b\bar{b}}, R_{\jet} ]}^\infty d R \, S_2(R)},
\end{equation}
where $\hat{R}_{b\bar{b}} = 2 \cdot 125\,\mathrm{GeV}/{p_{T,\jet}}$, $a = 0.75$, and $a'=1.25$.
For the upper boundary, we take the minimum between the boundary $a' \hat{R}_{b\bar{b}}$ and the jet radius $R_{\jet}$.
The angular scale beyond $R_{\jet}$ is mainly covered by large angle radiations rather than soft and collinear radiations from the $b$ partons.
The $b$ partons from the Higgs boson do not emit parton shower in large angle because the whole system is color neutral.
Therefore, we restrict the upper bound of the integral in the numerator up to $R_{\jet}$ while include the integral beyond the upper bound to the denominator.

\begin{figure}
\begin{center}
\includegraphics[width=0.48\textwidth]{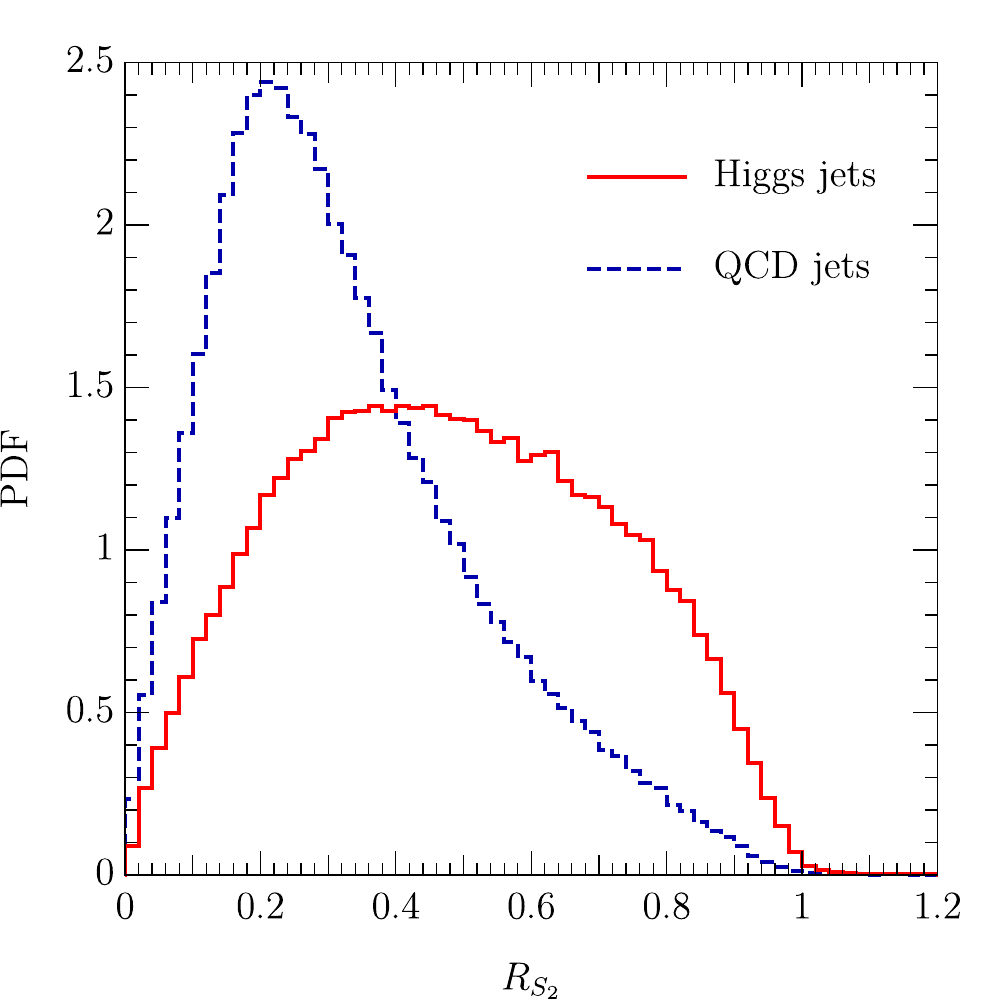}
\end{center}
\caption{
Histogram of $R_{S_2}$ of the Higgs jets and the QCD jets in the training sample.
The bin height is the probability distribution function (PDF) of the given bin.
}
\label{fig:Plot_RS2_dist}
\end{figure}

The probability that a QCD jet emits another hard parton at $\hat{R}_{b\bar{b}}$ is small; hence, this ratio works as a classifier.
Even if a QCD jet accidentally has substructure from parton splittings, the ratio of the momenta of two partons are different from that of Higgs jets.
It is typically small,
\begin{equation}
R_{S_2} 
= \frac{2 p_{T,i} p_{T,j} }{ p_{T,i}^2 + p_{T,j}^2} 
= \frac{2 z (1-z) }{1 - 2 z (1-z)},
\end{equation}
where $z$ is a fraction of momemtum of the one of the partons, i.e., $p_{T,i} = z p_{T,q/g}$ and $p_{T,j} = (1-z) p_{T,q/g}$. The fraction $z$ tends to be much smaller than 1  and suppresses the numerator. 
In contrast, the phase-space of the Higgs boson decay is  symmetric; hence, the two terms in \eqref{eqn:spectrum2_sig} are in a similar order and $R_{S_2}$ is approximately 1. 
Therefore, the typical Higgs jets and QCD jets have different $R_{S_2}$ values as in \figref{fig:Plot_RS2_dist} and the ratio works as a classifier.

To compare the performance of $R_{S_2}$ with other observables, we consider an observable $D_2$ \cite{Larkoski:2014gra} for the two-prong substurcture identification.  
The variable $D_2$ is defined by a ratio of two-point and three-point energy correlation functions $e_2^\beta$ and $e_3^\beta$ as follows,
\begin{eqnarray}
e_2^\beta
& = &
\frac{1}{p_{T,\jet}^2} \sum_{\substack{i,j \in \jet \\ i<j}} p_{T,i} p_{T,j} R_{ij}^\beta ,
\\
e_3^\beta
& = &
\frac{1}{p_{T,\jet}^3} \sum_{\substack{i,j,k \in \jet \\ i<j<k}} p_{T,i} p_{T,j} p_{T,k} R_{ij}^\beta R_{jk}^\beta R_{ki}^\beta ,
\\ 
D_2^\beta 
& = &
\frac{ e_3^\beta }{ (e_2^\beta)^3 } ,
\end{eqnarray}
where the summations of $e_2^\beta$ and $e_3^\beta$ run over all jet constituents. 
We consider angular exponents $\beta=0,5$, $1$, $2$ and $4$ for further discussion, but we focus on $\beta=2$ when we discuss analyses with a single $D_2$.
The Higgs jets have a small $D_2$ value because $e_3^\beta$ is suppressed by collinear and soft radiations while $e_2^\beta$ is large because the pairs of jet constituents with $R_{ij} \sim R_{b\bar{b}}$ dominate.
The QCD jets do not have such suppression, and hence, the Higgs jets and the QCD jets cover different regions of $D_2$.

\begin{figure}
\begin{center}
\includegraphics[width=0.48\textwidth]{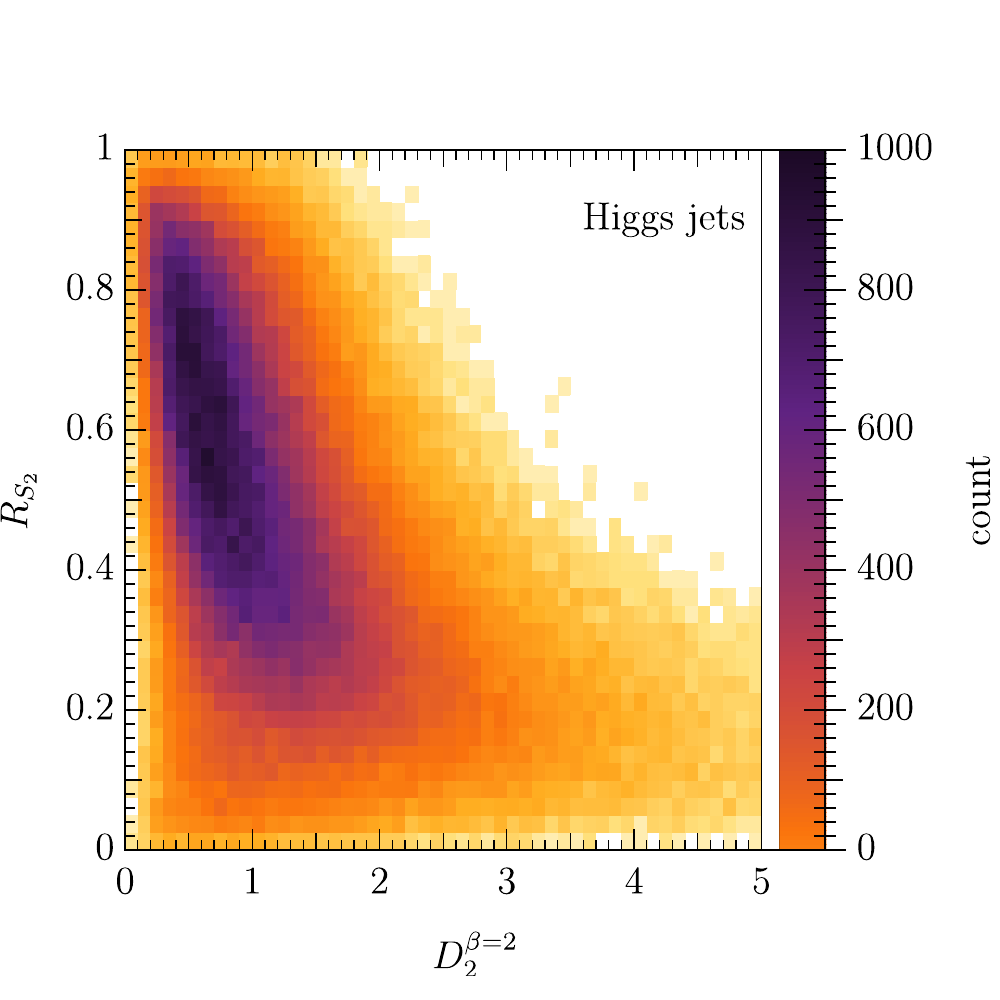}
\hfill	
\includegraphics[width=0.48\textwidth]{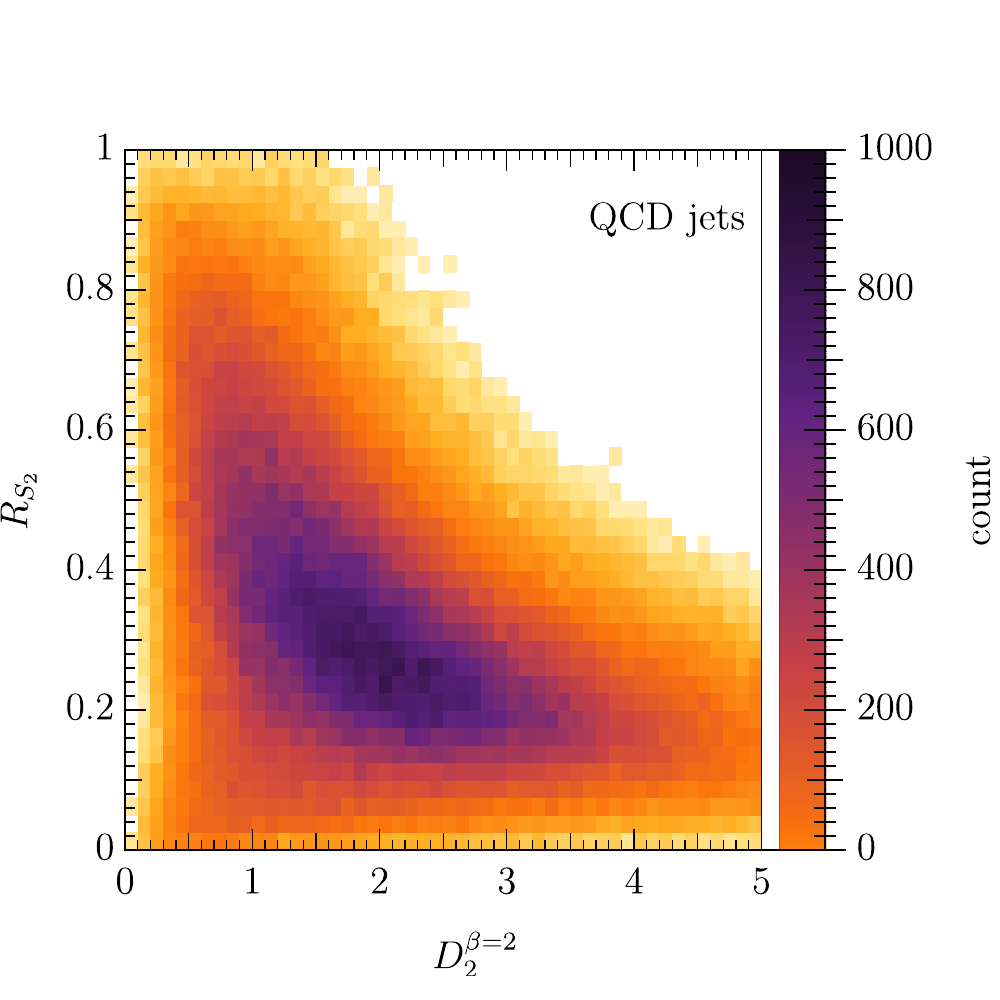}	
\end{center}
\caption{
Histograms of $(D_2^{\beta=2},R_{S_2})$ of the Higgs jets and the QCD jets in the training sample.
}
\label{fig:Plot_D2-RS2_dist}
\end{figure}

Since $R_{S_2}$ and $D_2^{\beta=2}$ are sensitive to the two-prong substructure, moderate correlation between them is expected.
We show histograms of $(D_2^{\beta=2},R_{S_2})$ of the training sample, in \figref{fig:Plot_D2-RS2_dist}.
The Higgs jets prefer small $D_2$ and large $R_{S_2}$, while the QCD jets prefer large $D_2$ and small $R_{S_2}$.
Hence, there is correlation.

\begin{figure}
\begin{center}
\includegraphics[width=0.55\textwidth]{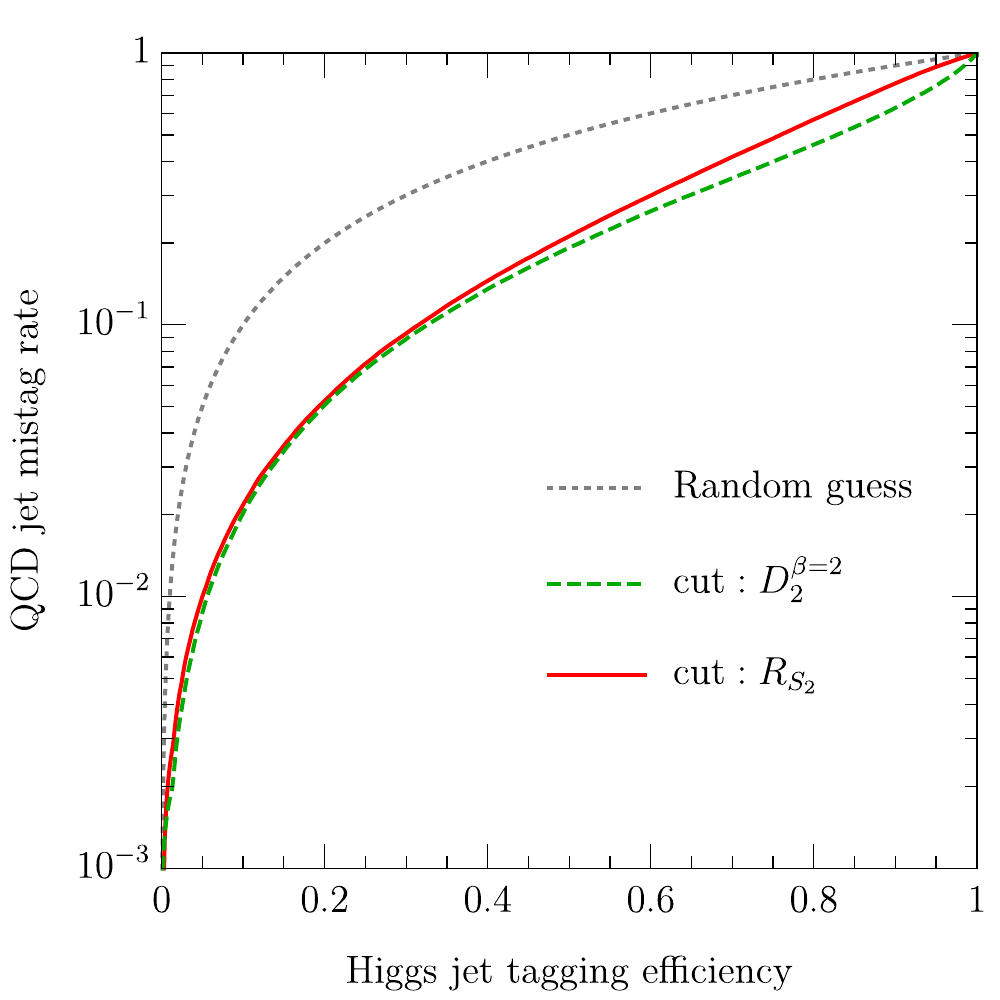}	
\end{center}
\caption{
ROC curves of $D_2$ and $R_{S_2}$ for classifying Higgs jets and QCD jets.
}
\label{fig:ROC_CUT}
\end{figure}

We show the receiver operating characteristic (ROC) curves of $R_{S_2}$ and $D_2^{\beta=2}$ in \figref{fig:ROC_CUT} to compare the classification performance. 
Since $R_{S_2}$ and $D_2^{\beta=2}$ are correlated, they show similar performance. 
At the Higgs tagging efficiency 0.2 (0.4), the QCD jet mistag rate is 0.0506 (0.135) for $D_2^{\beta=2}$ and 0.0525 (0.145) for $R_{S_2}$.

Note that $R_{S_2}$ is not a unique classifier. 
However, building the sophisticated variable from $S_2(R)$ is not the scope of this paper.
Instead, we will let the neural network build an optimized variable from $S_2(R)$ for the Higgs jet classification.

\section{Spectral Analysis with Artificial Neural Networks}
\label{sec:4}

We now feed the event-by-event binned spectra $S_2(R;\Delta R)$ to our ANN and build a neural network classifier between the Higgs jets and QCD jets.
First, we prepare an equal number of Higgs jets and QCD jets to avoid overfitting from unbalanced data.
We use \texttt{TFLearn} \cite{tflearn2016} with backend \texttt{TensorFlow} \cite{tensorflow2015-whitepaper} for the ANN analysis. 
An input set we consider includes $S_2(R;0.1)$ up to angular scale $R < 2$,
\begin{equation}
\{ x_i \}_{S_2} = \{ \ptj, \mj, S_2(0;0.1), \cdots, S_2(1.9;0.1) \}.
\label{eqn:input_s2}
\end{equation}
Note that $R=2$ is the diameter of our jet definition.
All the input data $\{x_i\}$ are standardized, i.e., $x_i \rightarrow ( x_i - \bar{x}_i ) / \sigma(x_i)$, where $\bar{x}_i$ and $\sigma(x_i)$ are the mean and the standard deviation of $x_i$ of the whole training sample including both Higgs jets and QCD jets. 
The network is configured with four hidden layers having $(400,300,200,100)$ nodes with the ReLU activation functions, $f(x) = \max(0,x)$, and an output layer with two nodes having the softmax activations which map inputs to a Higgs-like score $y_{S_2}$.
To avoid overtraining, we insert dropout layers \cite{JMLR:v15:srivastava14a} with rate 20\% between each hidden layer.
The network is trained by Adam optimizer \cite{ADAM} with learning rate 0.001, $\beta_1=0.99$ and $\beta_2=0.999$
minimizing a categorical cross-entropy as a loss function,
\begin{equation}
L =  \frac{1}{N_{\mathrm{events}}} \sum_{ \mathrm{events}}
\begin{cases}
- \log [y_{S_2}(\{x_i\})] & \mathrm{Higgs\,jets},  \\
- \log [ 1-y_{S_2}(\{x_i\}) ] & \mathrm{QCD\,jets},
\end{cases}
\end{equation}
where $N_{\mathrm{events}}$ is the number of events in the training event set.
We call this network as $\ANN{S_2}$.
In the trained network, Higgs jets have scores near 1, while QCD jets have scores near 0.
We validate $\ANN{S_2}$ using the testing samples.

We compare the performance of $\ANN{S_2}$ with that of a network trained with $D_2$.
We prepare another neural network which maps following inputs to the Higgs-like score $y_{D_2}$,
\begin{equation}
\{ x_i \}_{D_2} = \{ \ptj, \mj, D_2^{\beta=2} \} .
\label{eqn:input_d2}
\end{equation}
Again, the input data is normalized to $[0,1]$, i.e., $x_i \rightarrow ( x_i - \min x_i) / (\max x_i - \min x_i)$, where $\max x_i$ and $\min x_i$ are the maximum and the minimum of $x_i$ in the training sample respectively.
We use smaller hidden layers $(100,100)$ ReLU nodes because the number of inputs is smaller.
The other ANN setups are identical to the $S_2(R)$ analysis.
We call this network as $\ANN{D_2}$.

In our approach, we do not use individual pixels as inputs; therefore, the binned spectrum is affected by both soft and hard calorimeter activities.
To make ANN learn a hierarchy between soft and hard radiations, variables after jet trimming \cite{Krohn:2009th} are useful.
To obtain trimmed quantities, we first reconstruct $k_T$ subjets \cite{Catani:1993hr,Ellis:1993tq} with $R_{\mathrm{sub}} = 0.2$ from constituents of the jet  and remove subjets having transverse momentum $p_{T,\mathrm{subjet}} < f_{\mathrm{cut}} \cdot \ptj$, where $f_{\mathrm{cut}} = 0.05$.
In the right panel of \figref{fig:jet_profile_char}, 
we show typical two-point correlation spectra of the trimmed jet constituents, $S_{2,tr}(R)$, of a Higgs jet and a QCD jet.
The spectra $S_2(R)$ before trimming are shown in the central panel.
The two-prong substructure of a Higgs jet is consisted by hard activities, and the double peak structure appears both in $S_2(R)$ and $S_{2,tr}(R)$.
On the other hand, the spectrum of a QCD jet is significantly changed after trimming, which means that soft activities dominate the spectrum $S_2(R)$.
This shows a difference between $S_2(R)$ and $S_{2,tr}(R)$ contains useful information for the classification.

We then prepare three networks $\ANN{D_{2}+tr}$, $\ANN{D_{2}^\beta+tr}$ and $\ANN{S_{2}+tr}$ taking inputs $\{ x_i \}_{D_{2}+tr}$, $\{ x_i \}_{D_{2}^\beta+tr}$ and $\{ x_i \}_{S_{2}+tr}$ respectively,
\begin{eqnarray}
\{ x_i \}_{D_{2}+tr} 
& = &
\{ x_i \}_{D_{2}} \cup \{ p_{T,\jet,tr}, m_{\jet,tr} , D_{2,tr}^{\beta=2} \} ,
\label{eqn:input_d2_tr}
\\
\{ x_i \}_{D_{2}^\beta+tr} 
& = &
\{ x_i \}_{D_{2}+tr} \cup \{ 
 D_{2}^{\beta=0.5},
 D_{2}^{\beta=1},
 D_{2}^{\beta=2},
 D_{2}^{\beta=4},
 D_{2,tr}^{\beta=0.5},
 D_{2,tr}^{\beta=1},
 D_{2,tr}^{\beta=2},
 D_{2,tr}^{\beta=4}
 \} ,
\label{eqn:input_d2beta_tr}
\\
\{ x_i \}_{S_{2}+tr} 
& = &
\{ x_i \}_{S_{2}} \cup \{ p_{T,\jet,tr}, m_{\jet,tr} , S_{2,tr}(0.0;0.1), \cdots, S_{2,tr}(1.9;0.1) \},
\label{eqn:input_s2_tr}
\end{eqnarray}
where the variables calculated after trimming have a subscript $tr$.
The other ANN setups are the same as that of $\ANN{D_{2}}$ and $\ANN{S_{2}}$. 
The networks $\ANN{D_{2}+tr}$, $\ANN{D_{2}^\beta+tr}$ and $\ANN{S_{2}+tr}$ give us Higgs-like scores $y_{D_{2}+tr}$, $y_{D_{2}^\beta+tr}$ and $y_{S_{2}+tr}$.
Note that $\ANN{D_{2}^\beta+tr}$ takes the $D_2^\beta$ with various angular exponents $\beta = 0.5$, $1$, $2$, and $4$ as inputs.

\begin{figure}
\begin{center}
\includegraphics[width=0.55\textwidth]{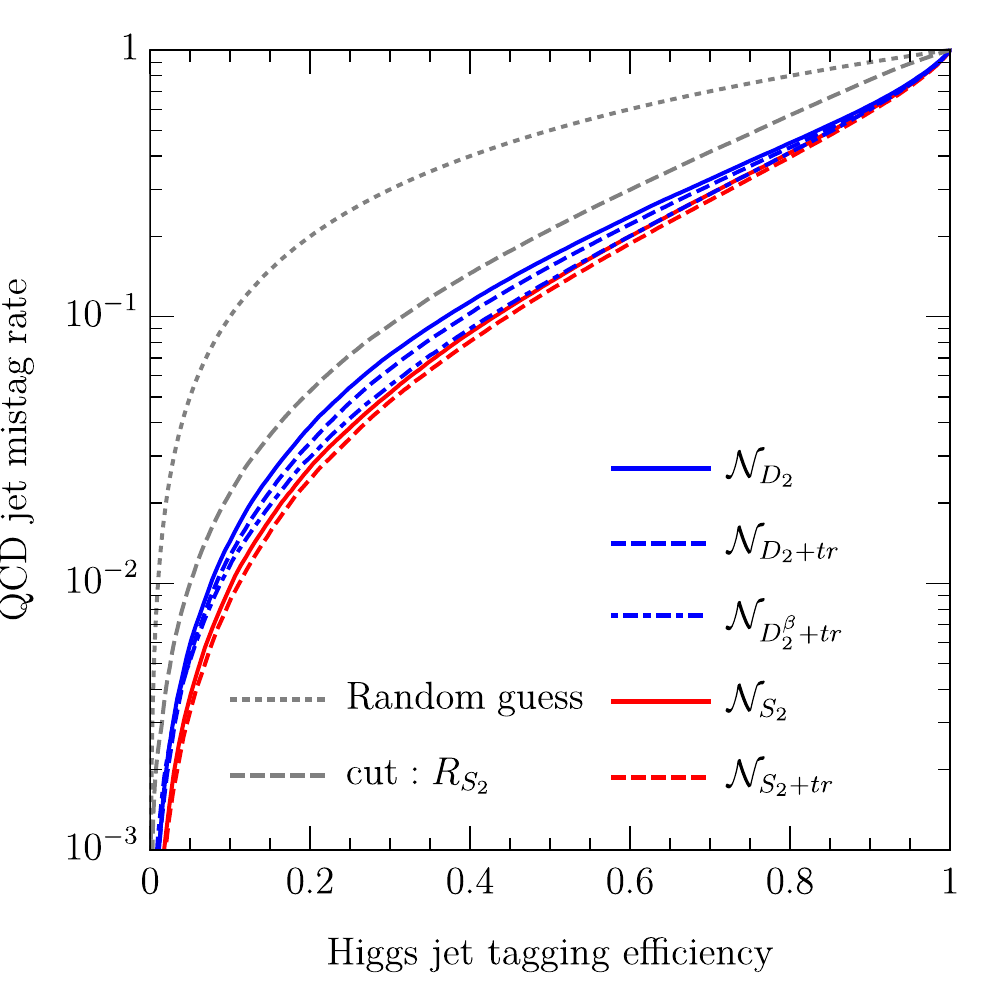}	
\end{center}
\caption{
ROC curves of ANN classifiers for Higgs jets and QCD jets with inputs \eqref{eqn:input_s2}, \eqref{eqn:input_d2}, \eqref{eqn:input_d2_tr}, 
\eqref{eqn:input_d2beta_tr}, and \eqref{eqn:input_s2_tr}. We overlay the cut-based ROC curves of $R_{S_2}$ shown in \figref{fig:ROC_CUT} as a gray dashed line for a reference.
}
\label{fig:ROC_S2vsD2}
\end{figure}

To compare the information contained in $D_2$ and $S_2(R)$, we show the ROC curves of our ANN analyses with $\ANN{D_{2}+tr}$, $\ANN{D_{2}+tr}$, $\ANN{D_{2}^\beta+tr}$, $\ANN{S_{2}}$, and $\ANN{S_{2}+tr}$ in \figref{fig:ROC_S2vsD2}.
The ROC curves show that the ANN with $S_2(R;\Delta R)$ rejects more QCD jets for a fixed Higgs jet efficiency.
At the Higgs tagging efficiency 0.4 (0.2), the QCD jet mistag rate of $\ANN{S_{2}+tr}$, which shows the best performance among the classifiers, is 0.0246 (0.0807).
The mistag rate is reduced by 51.4\% (40.2\%) compared to that of the cut-based analysis of $R_{S_2}$ while the mistag rate is still reduced by 17.1\% (10.5\%) compared to that of $\ANN{D_{2}^{\beta}+tr}$.
This is expected because $\ANN{S_{2}+tr}$ uses two-point energy correlation from $S_2(R;\Delta R)$ and infers three-point energy correlation from correlations between different angular scales.
For example, $S_2(R;\Delta R)$ of a three-prong jet having angular scales $R_1$, $R_2$, and $R_3$ has three peaks away from $R=0$. The intensity of each peak gives a three-point energy correlation function, $e_3^\beta \approx \ptj^{-1} \cdot \sqrt{ (\Delta R)^3 S_2(R_1;\Delta R) S_2(R_2;\Delta R) S_2(R_3;\Delta R)} R_1^\beta R_2^\beta R_3^\beta $.
Hence, $\ANN{S_{2}}$ and $\ANN{S_{2}+tr}$ have better discrimation power than  $\ANN{D_{2}+tr}$, $\ANN{D_{2}+tr}$ and $\ANN{D_{2}^\beta+tr}$.
Also, adding trimmed observables allows ANN to learn hard and soft substructures separately.
Hence, the ANN solve degeneracy in the variables before trimming and reject QCD jets better.
One interesting feature is that $\ANN{D_{2}^{\beta}+tr}$ suppress QCD jets about as equal as $\ANN{S_2}$ in the region of high tagging efficiency, but the difference between $\ANN{D_{2}^{\beta}+tr}$ and $\ANN{S_2}$ is large in the region of low tagging efficiency.
This gap in the ROC curves implies that $S_2(R)$ has additional information, which we will be discussed in the later part.

Note that the ANN inputs include $p_{T,\jet}$; as a result, the ANN partially use the jet $p_T$ distribution of Higgs jets and QCD jets for the classification.
To remove the $p_T$ dependence, we may resample the training set whose signal and background $p_T$ distribution is same, or use a weighted loss function for an imbalanced training set, or train adversarial neural networks \cite{Louppe:2016ylz,Shimmin:2017mfk}.
For the comparison of the ANN's with different inputs, we do not need to use these techniques because $p_{T,\jet}$ is a common input for all the ANN analyses.
We have to pay attention to the bias from $p_{T,\jet}$ distribution of signal and background when this analysis is applied to the experimental data.

\begin{figure}
\begin{center}
\begin{tabular}{cc}
\includegraphics[width=0.49\textwidth]{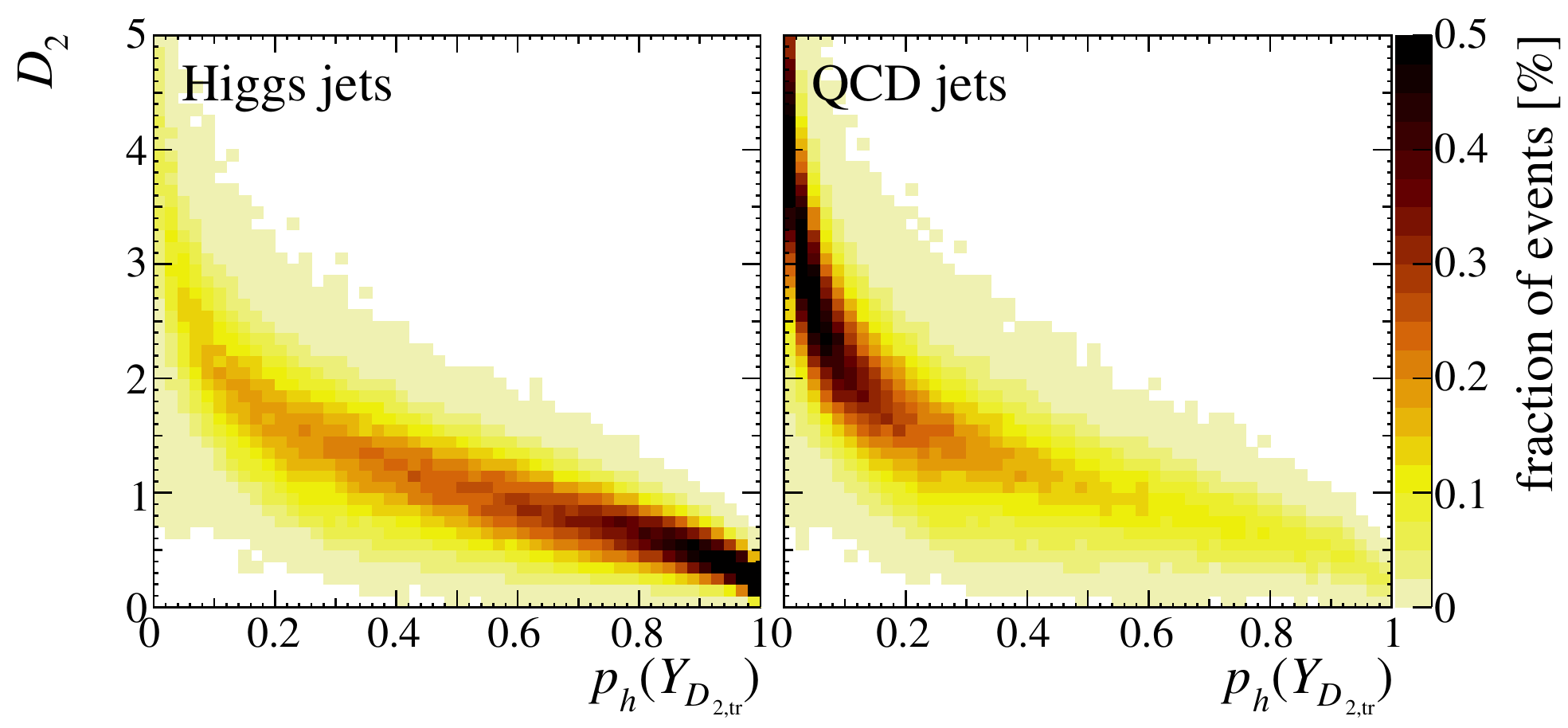} &
\includegraphics[width=0.49\textwidth]{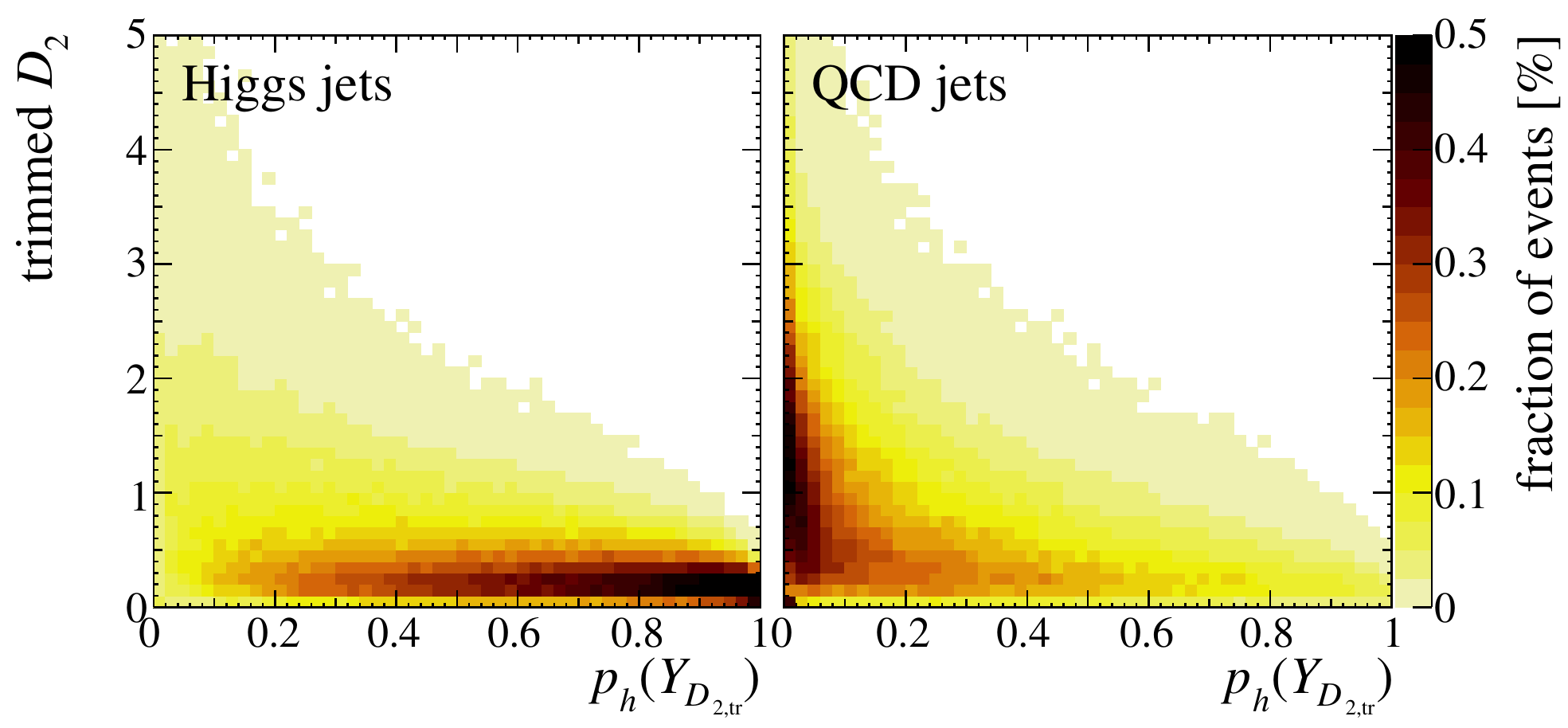}\\
\includegraphics[width=0.49\textwidth]{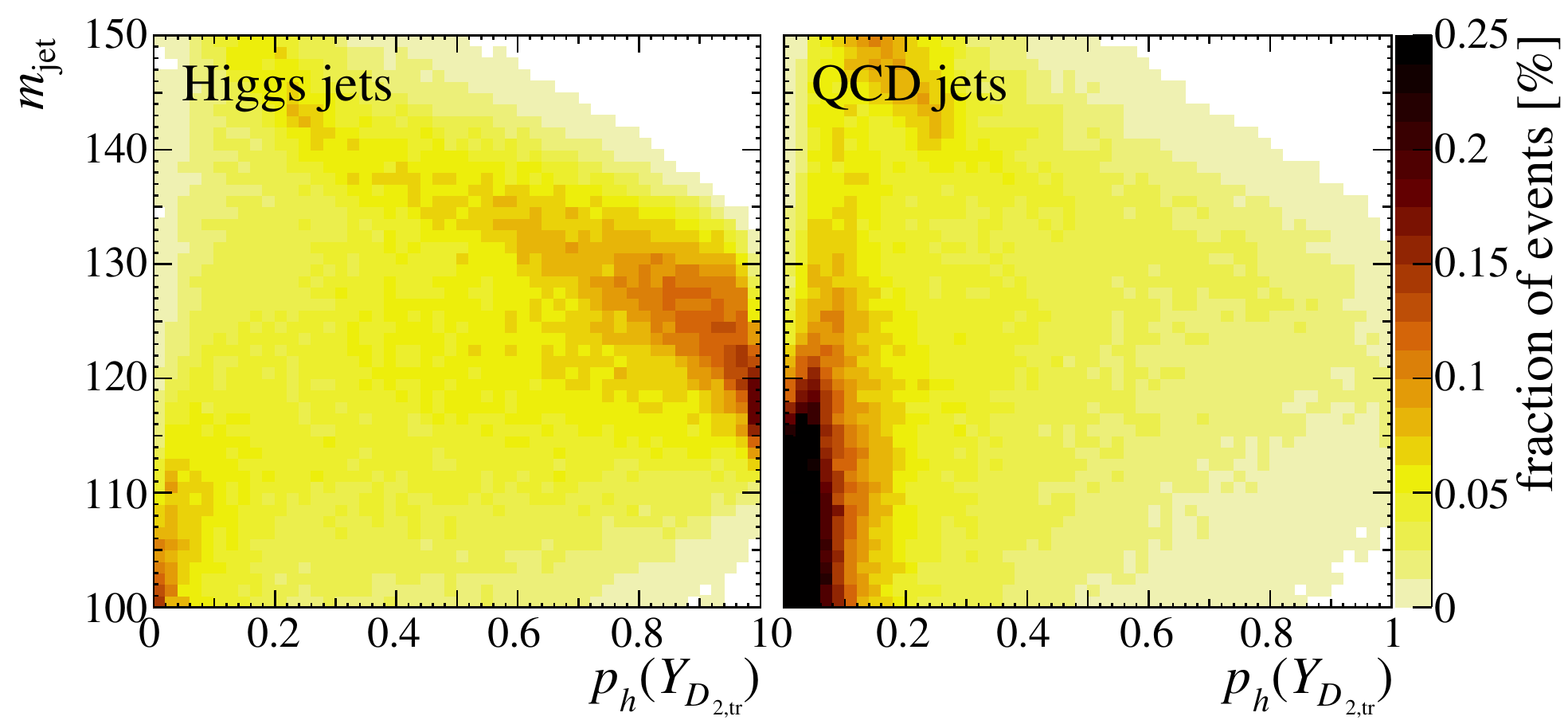} &
\includegraphics[width=0.49\textwidth]{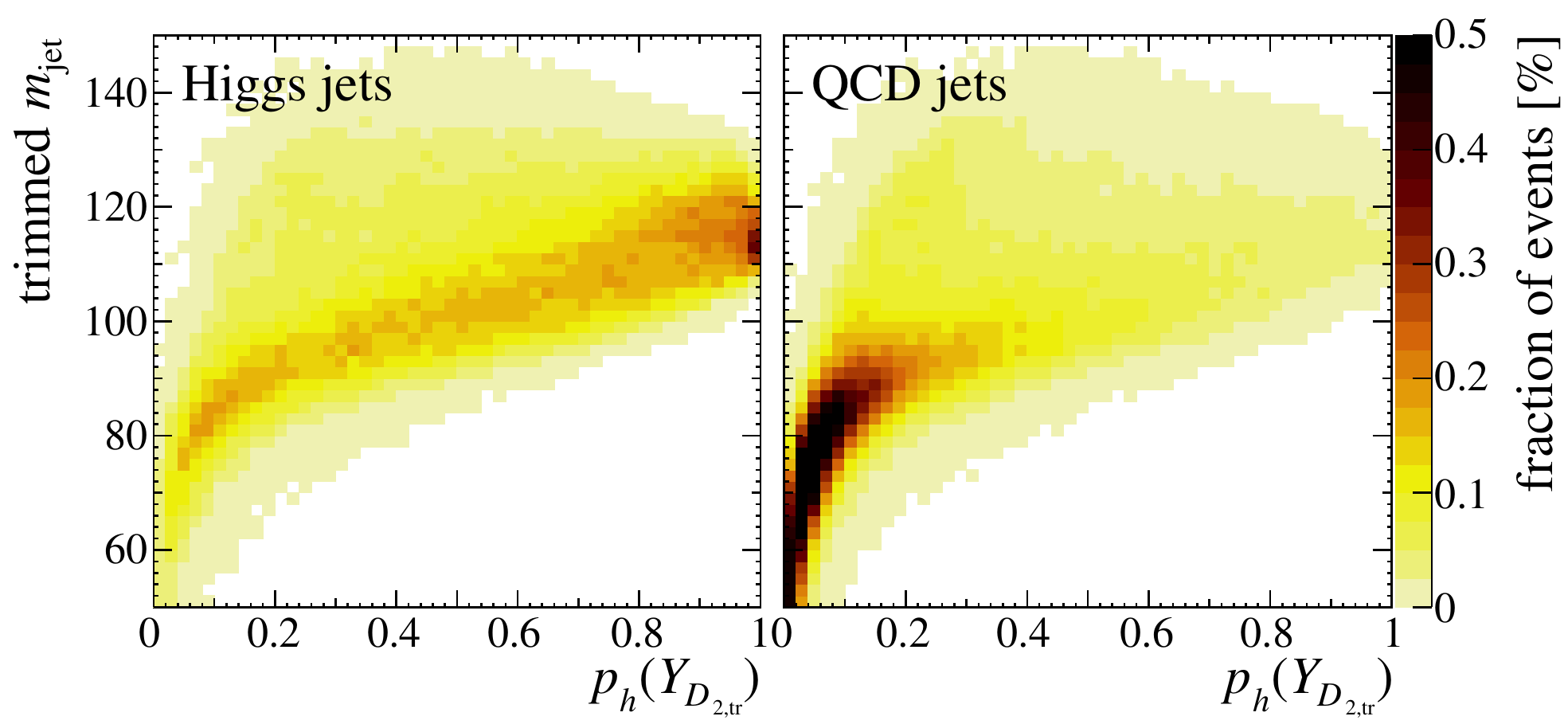}\\
\includegraphics[width=0.49\textwidth]{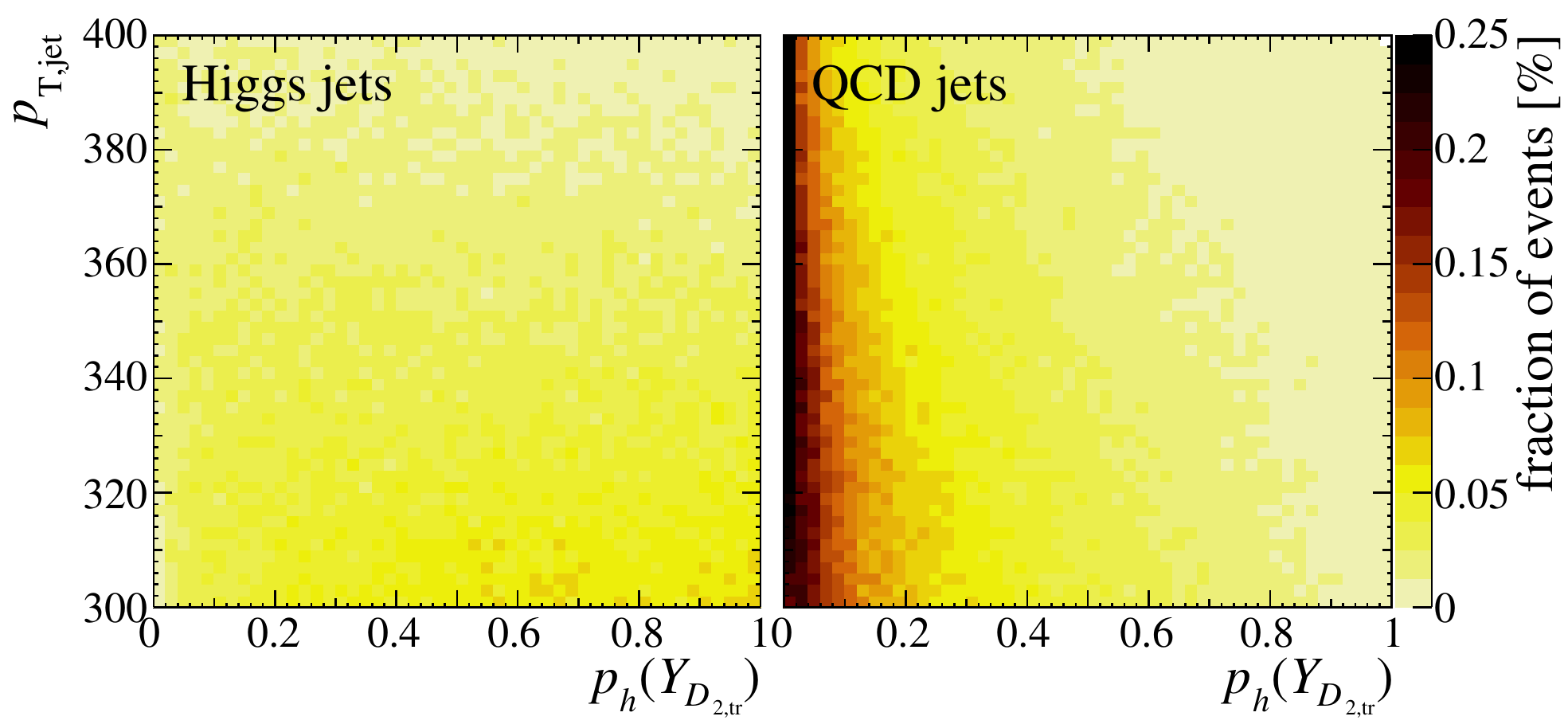} &
\includegraphics[width=0.49\textwidth]{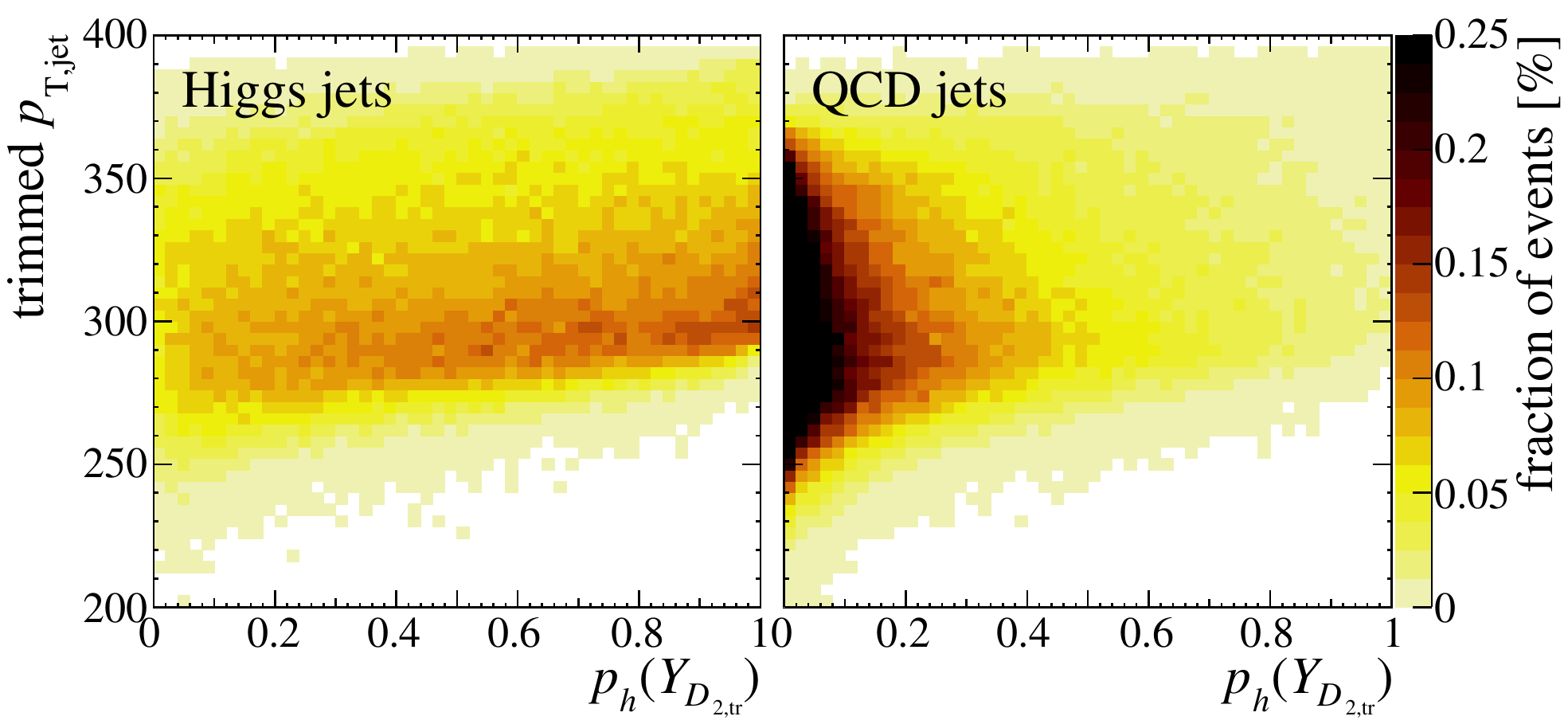}
\end{tabular}
\end{center}
\caption{
The distributions of Higgs jets and QCD jets in the Higgs-like probability $p_h(Y_{D_2+tr})$ and inputs in $\{x_i\}_{D_2+tr}$ planes.
For each pair of figures, a left panel is the distribution of Higgs jets and a right panel is the distribution of QCD jets.
}
\label{fig:D2_performance}
\end{figure}

The network $\ANN{D_2+tr}$ uses a small number of inputs; therefore, it is easy to check the ANN outputs of the input parameters $\mj$, $\ptj$, and $D_2$.
We show distributions of the events in the Higgs-like probability $p_h ( Y_{D_2+tr} )$ and one of the inputs $\{x_i\}_{D_2+tr}$ plane in \figref{fig:D2_performance}. 
The Higgs-like probability of $\ANN{X}$, $p_h ( Y_{X} )$, is defined by the probability of getting a score $y_{X}$ less than $Y_{X}$ in the Higgs jet samples,
\begin{equation}
p_h ( Y_{X} ) = \mathrm{Pr}( y_{X} < Y_{X} \,|\, \mathrm{Higgs}\;\mathrm{jets}),
\label{eqn:prob_higgs-like}
\end{equation}
where $\mathrm{Pr}(C|H)$ represents the conditional probability of $C$ given $H$.
A large $p_h ( Y_{X})$ means that the given jet is more Higgs-like in $\ANN{X}$.
\figref{fig:D2_performance} shows that $\ANN{D_2+tr}$ tries to select jets having $\mj$ around 125 GeV up to energy loss, and small $D_2$ for capturing two-prong jets.
Note that the $\ANN{D_2+tr}$ is trained by the QCD jets and Higgs jets with $m_h = 125\;\mathrm{GeV}$.
The $m_\jet$ and trimmed $m_\jet$ of Higgs jet is tend to be higher (lower) than the mass of Higgs boson and a jet with $m_\jet \sim 120\,\mathrm{GeV}$ is regarded most likely as a Higgs jet.
These shifts of the mass from the input Higgs mass are due to contamination of other activities, or large angle radiations from $b$ jets.
The mistagged QCD jets with high $p_h(Y_{D_2+tr})$ cover similar phase-space, where $D_{2,tr}$ is small, $m_{\jet,tr} \sim 115\;\mathrm{GeV}$, and $\mj \sim 125\;\mathrm{GeV}$.
Meanwhile, the $\mj$ and $m_{\jet,tr}$ distributions of QCD jets in $p_h(Y_{D_2+tr}) < 0.2$ and $p_h(Y_{D_2+tr}) < 0.3$ in \figref{fig:D2_performance} show relatively low probability for $\mj \sim 130\;\mathrm{GeV}$ and $m_{\jet,tr} > 100\;\mathrm{GeV}$, repsectively.
This means, $\ANN{D_2+tr}$ labels events in the particular mass window as Higgs jet; hence, $\ANN{D_2+tr}$ rejects QCD jets in a similar fashion that a cut-based approach rejects QCD events.

\begin{figure}
\begin{center}
\includegraphics[width=0.65\textwidth]{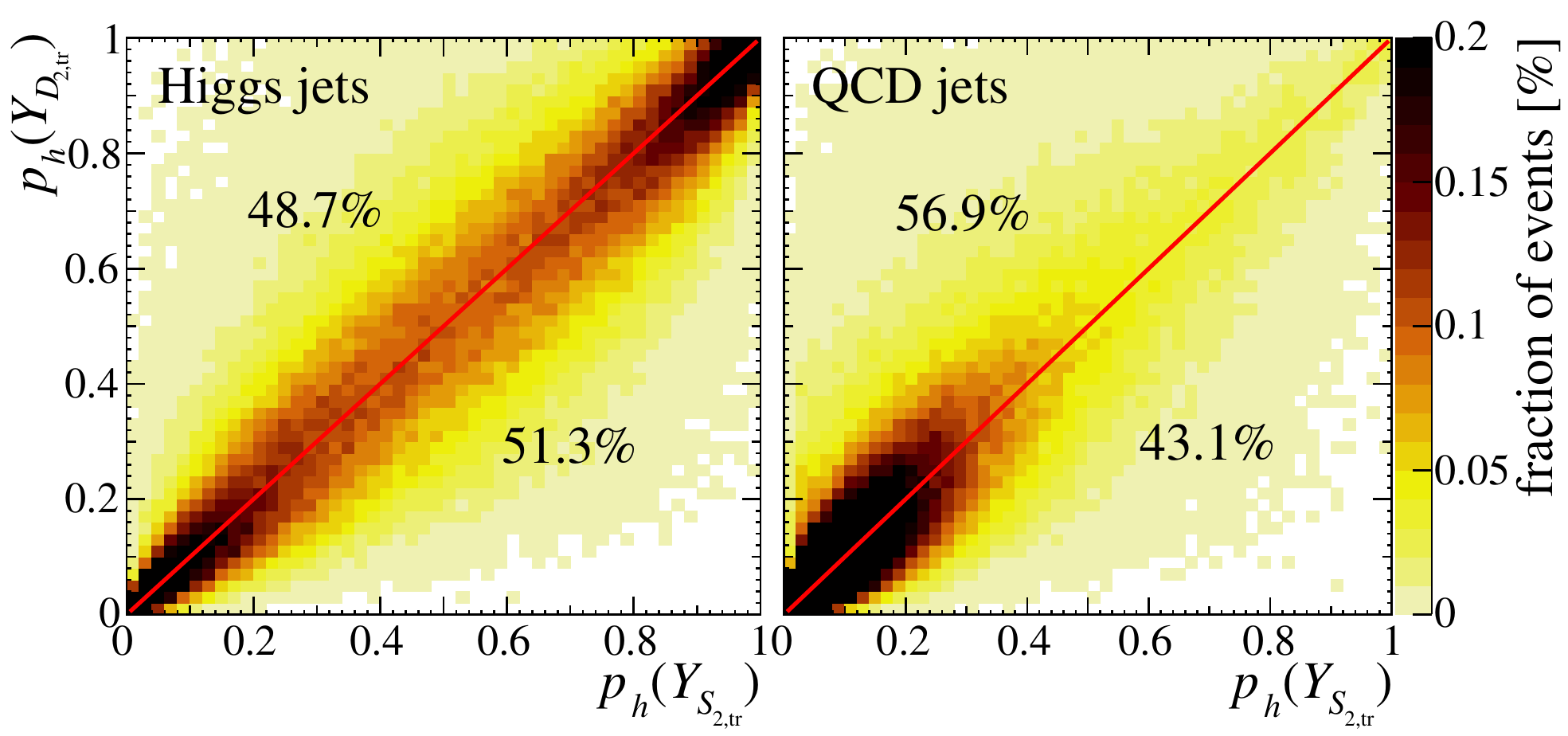}
\end{center}
\caption{
The distributions of Higgs-like probability in $p_h (Y_{S_2+tr})$ and $p_h (Y_{D_2+tr})$ planes.
The numbers in upper left and lower right area represent the fraction of events in the area.
}
\label{fig:ROC_S2vsD2_score_corr}
\end{figure}

We compare the Higgs-like probabilities in $\ANN{D_2+tr}$ and $\ANN{S_2+tr}$ in \figref{fig:ROC_S2vsD2_score_corr} to find out the origin of improvement.
The events are widely spreading around the line $p_h ( Y_{D_{2}+tr} ) = p_h ( Y_{S_{2}+tr} )$, which means that those two analyses have different selection criteria.
To quantify the residual of the anticorrelation of $\ANN{D_2+tr}$ and $\ANN{S_2+tr}$, we show the fractions of the events in the upper triangular region $p_h ( Y_{D_{2}+tr} ) > p_h ( Y_{S_{2}+tr} )$ and the lower triangular region $p_h ( Y_{D_{2}+tr} ) < p_h ( Y_{S_{2}+tr} )$.
For Higgs jets, the lower triangular region contains more events compared with the upper triangular region, 51.3\% of the total events.
For QCD jets, the lower triangular region contains less events, 43.1\%.
Hence, $\ANN{S_2+tr}$ improves signal and background ratio $S/B$ from $\ANN{D_2+tr}$.

\begin{figure}
\begin{center}
\includegraphics[width=0.3\textwidth]{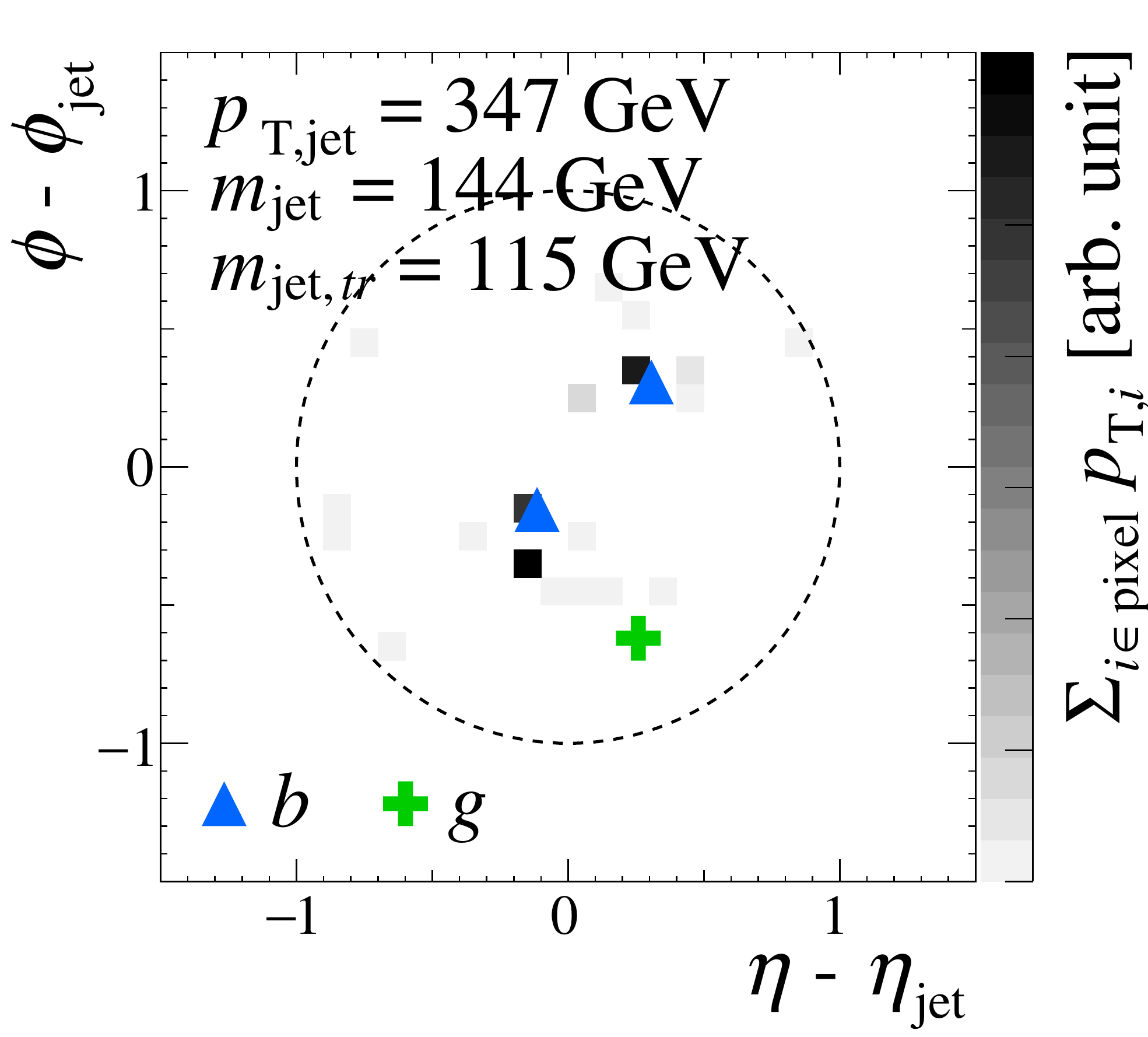}	
\includegraphics[width=0.3\textwidth]{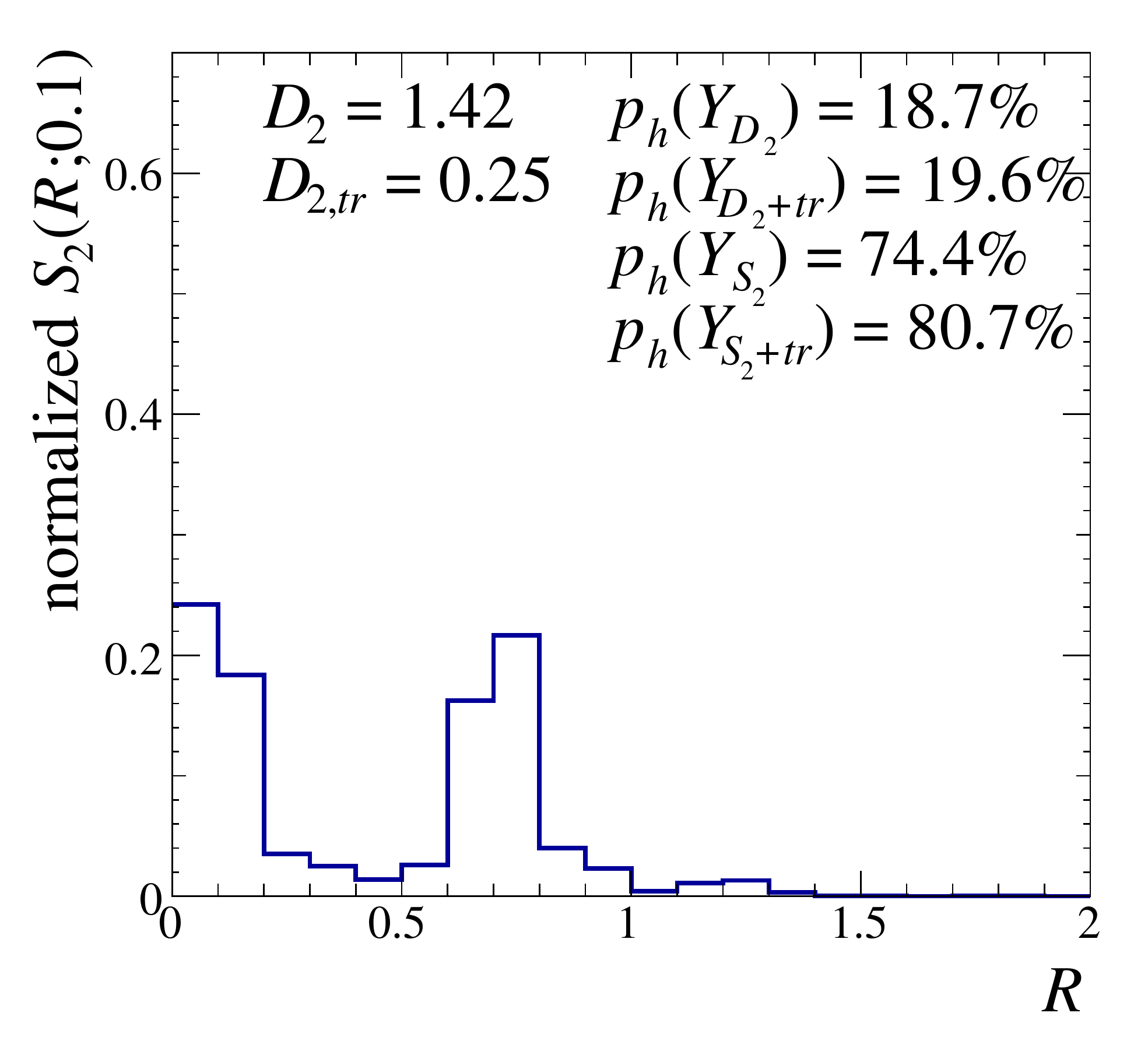}
\end{center}
\caption{
The pixelated jet image (left), and two-point spectrum $S_2(R;0.1)$ (right) of a Higgs jet. 
The event is Higgs-like in the neural networks with $S_2(R)$, $\ANN{S_2}$ and $\ANN{S_2+tr}$, but it is not in those with $D_2$, $\ANN{D_2}$ and $\ANN{D_2+tr}$.
The sum of the bins of the spectrum is normalized to 1. 
The blue triangles and green plus symbols in the pixelated jet images indicate direction of $b$ quarks and gluons obtained from the matrix element including order $\alpha_S$ radiations.
We show the Higgs-like probability $p_h(Y_X)$ in an ANN model $\ANN{X}$ defined by \eqref{eqn:prob_higgs-like}.
}
\label{fig:jet_profile_HighS2_LowD2}
\end{figure}

\begin{figure}
\begin{center}
\includegraphics[width=0.65\textwidth]{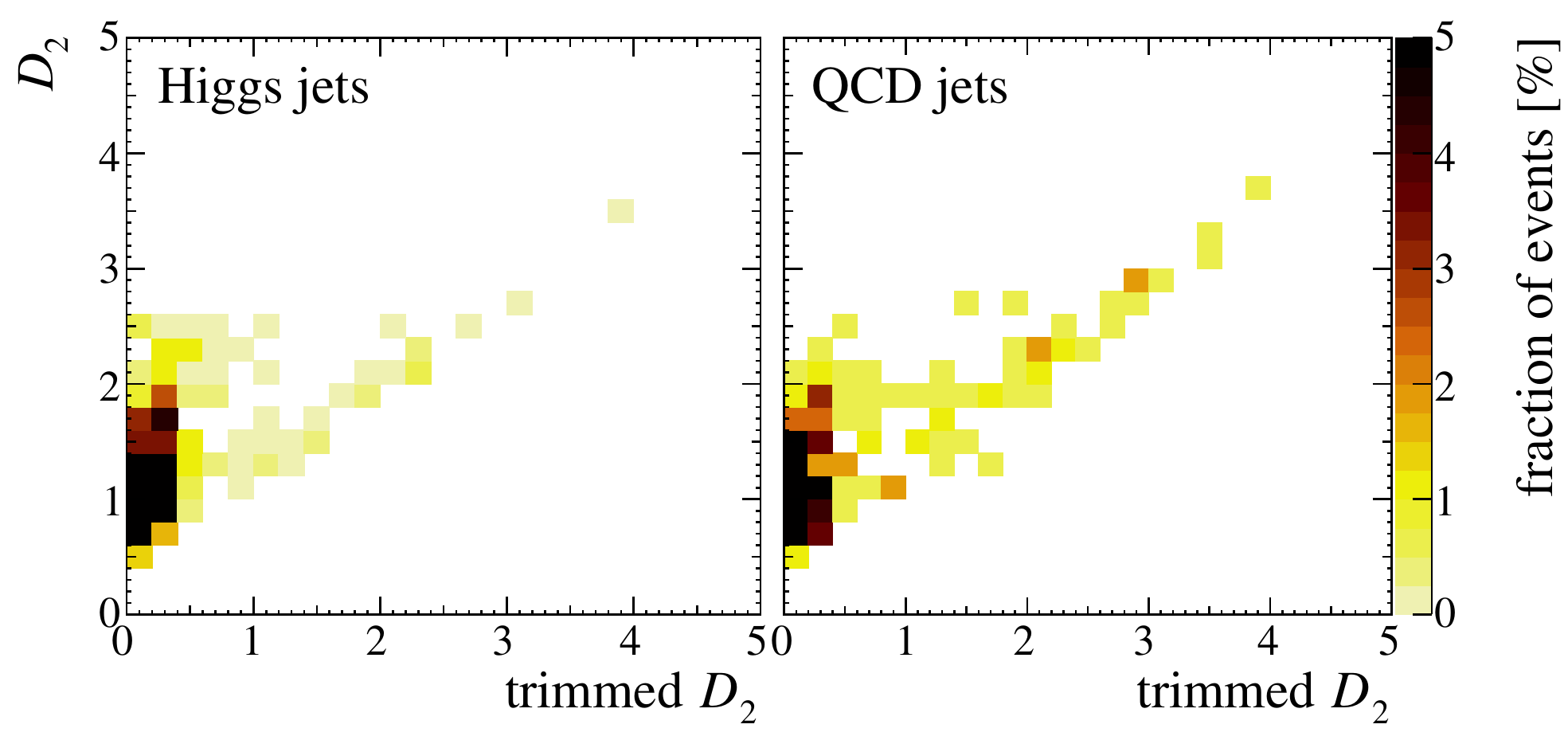}
\end{center}
\caption{
Distributions of Higgs jets (left) and QCD jets (right) in $D_2$ and $D_{2,tr}$, classified as Higgs jets in $\ANN{S_{2}+tr}$ but QCD jets in $\ANN{D_{2}+tr}$. 
We select events with Higgs-like probabilities $p_h ( Y_{D_{2,tr}} ) < 30\%$ and $p_h ( Y_{S_{2,tr}} )> 70\%$.
}
\label{fig:obs_HighS2_LowD2}
\end{figure}

To figure out how $\ANN{S_2+tr}$ accepts more Higgs jets while rejecting more QCD jets compared to $\ANN{D_2+tr}$, we will show three examples of the events located on the off-diagonal regions in \figref{fig:ROC_S2vsD2_score_corr}.
We show a Higgs jet in \figref{fig:jet_profile_HighS2_LowD2}, which is Higgs-like in $\ANN{S_2+tr}$ but regarded as a QCD jet in $\ANN{D_2+tr}$, $p_h(Y_{D_2+tr}) = 19.6\%$ and $p_h(Y_{S_2+tr}) = 80.7\%$. 
This jet has a moderate wide-angle radiation on top of two-prong substructure which increases $D_2$ significantly.
Remind that a Higgs jet originates from a color singlet particle while a QCD jet originates from a colored parton.
Such wide-angle radiation is easily generated from a colored parton compared to a color singlet particle.
$\ANN{D_2+tr}$ is distracted by a large $D_2$ and assigns this jet as a QCD jet even though the jet has small trimmed $D_2$. 
$\ANN{S_2+tr}$ must have determined the jet as Higgs-like from the information of microscopic radiation patterns in $S_2(R)$ which shows a clear double peak structure.
\figref{fig:obs_HighS2_LowD2} shows $D_2$ and $D_{2,tr}$ distributions in events having $p_h ( Y_{D_{2,tr}} ) < 30\%$ and $p_h ( Y_{S_{2,tr}} )> 70\%$.
We can see that some, but not all, events with large $D_2$ but small $D_{2,tr}$ fall into this region.

\begin{figure}
\begin{center}
\includegraphics[width=0.3\textwidth]{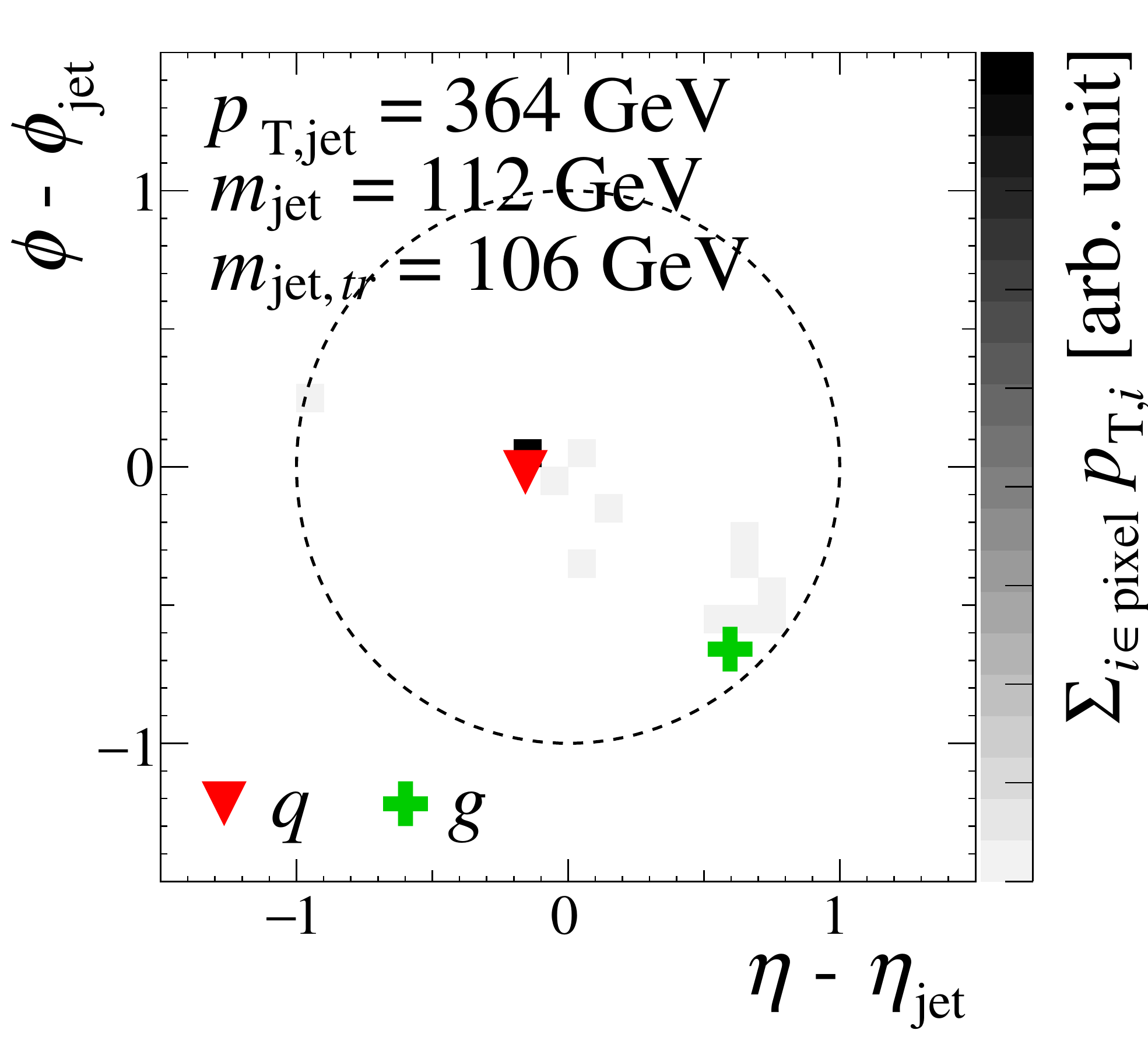}	
\includegraphics[width=0.3\textwidth]{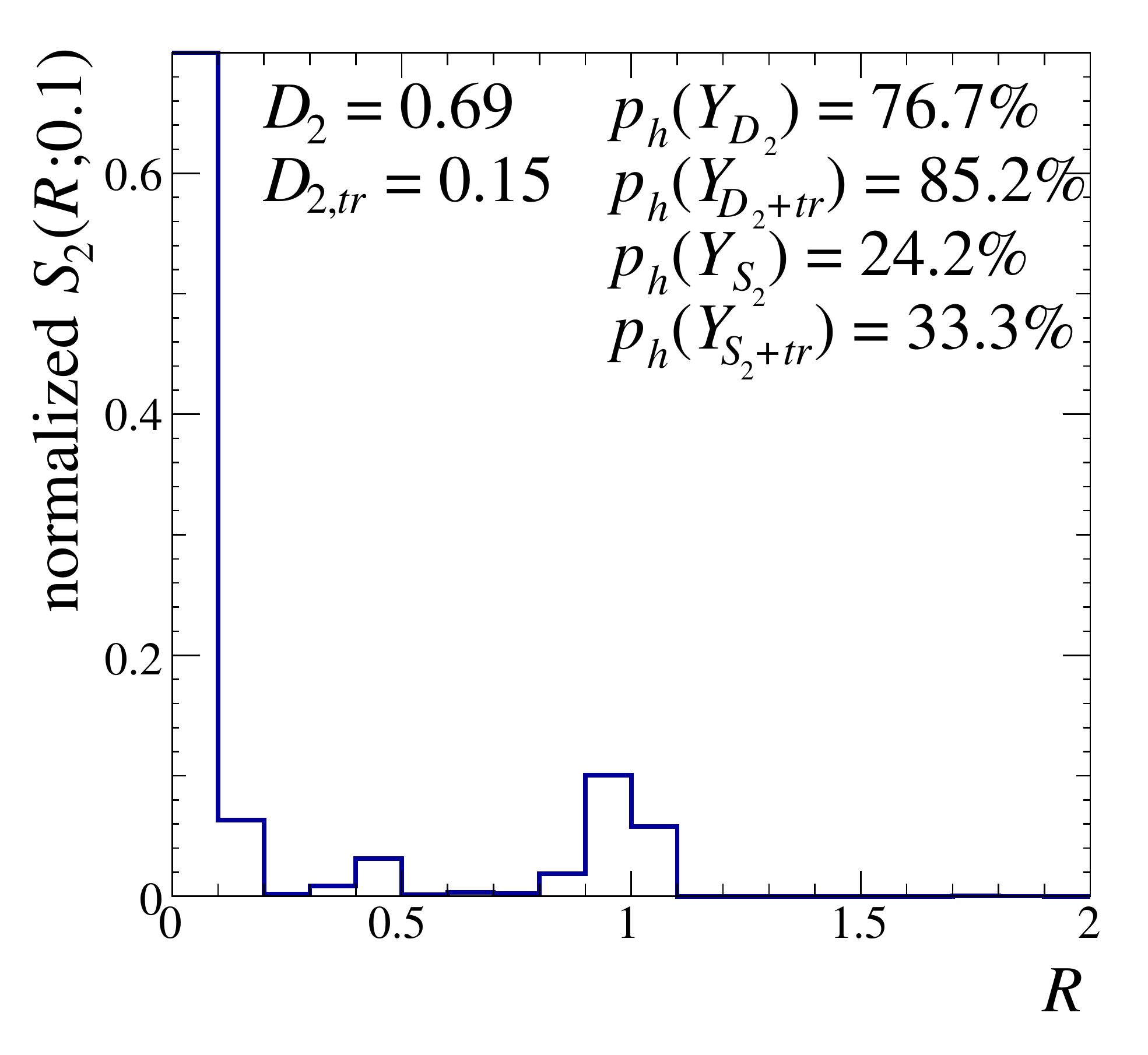}	
\end{center}
\caption{
The Pixelated jet image (left), and two-point spectrum $S_2(R;0.1)$ (right) of a Higgs jet. 
The event is Higgs-like in the neural networks with $D_2$, $\ANN{D_2}$ and $\ANN{D_2+tr}$, but it is not in those with $S_2(R)$, $\ANN{S_2}$ and $\ANN{S_2+tr}$.
The sum of the bins of the spectrum is normalized to 1. 
The red triangles and green plus symbols in the pixelated jet images indicate direction of light quarks and gluons obtained from the matrix element including order $\alpha_S$ radiations.
We show the Higgs-like probability $p_h(Y_X)$ in an ANN model $\ANN{X}$ defined by \eqref{eqn:prob_higgs-like}.
}
\label{fig:jet_profile_HighD2_LowS2}
\end{figure}

The second example in \figref{fig:jet_profile_HighD2_LowS2} is a jet classified as a Higgs jet in $\ANN{D_2+tr}$ but categorized as a QCD jet in $\ANN{S_2+tr}$.
This jet has evident two-prong substructure, and hence, $\ANN{D_2+tr}$ classifies this jet as a Higgs jet.
However, the two subjets are asymmetric in $p_T$.
Such events appear frequently in QCD jet samples.
We did not give subjet momenta to $\ANN{D_2+tr}$, and the ANN classify the jet as a Higgs jet.
In contrast, $S_2(R)$ knows the $p_T$ asymmetry by comparing the peak intensity; see \eqref{eqn:spectrum2_sig}.
Hence, $\ANN{S_2+tr}$ avoids these $p_T$ asymmetric events which appear often among QCD jets while $\ANN{S_2+tr}$ finds the cut on the subjet $p_T$ from the training samples.
In the mass-drop tagger \cite{Butterworth:2008iy}, the events with asymmetric $p_T$ subjets are removed by a cut.

\begin{figure}
\begin{center}
\includegraphics[width=0.3\textwidth]{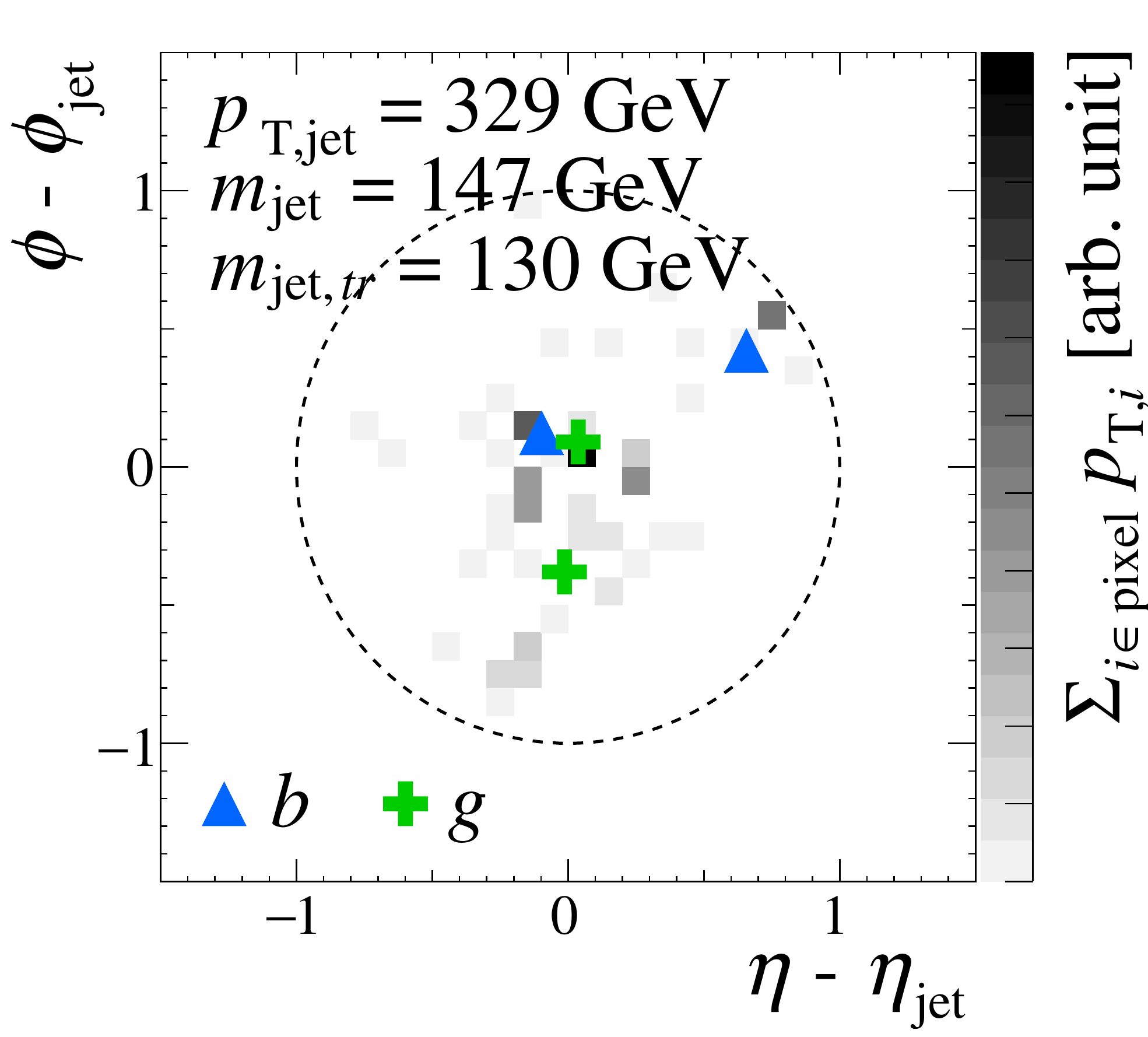}	
\includegraphics[width=0.3\textwidth]{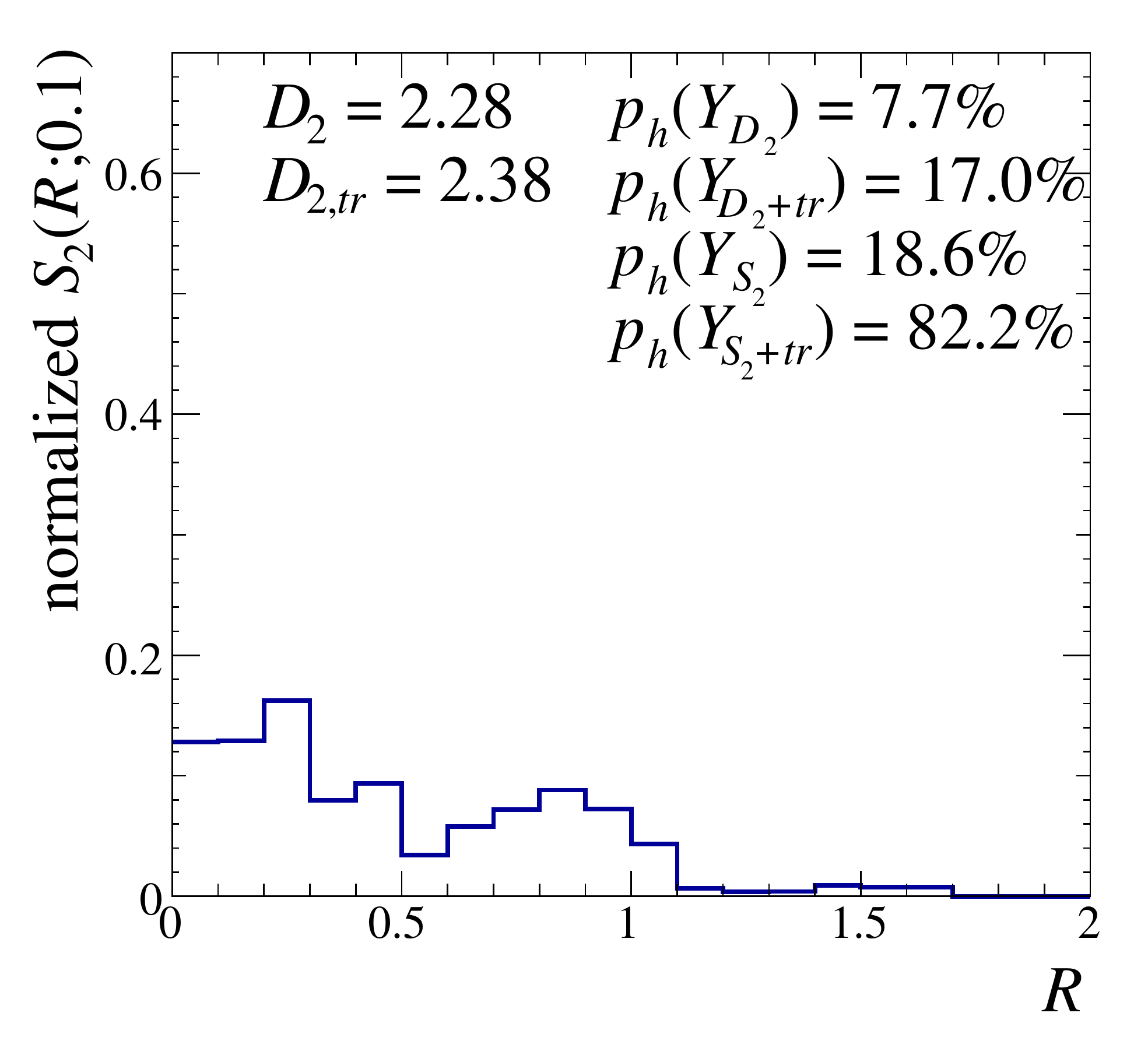}	
\includegraphics[width=0.3\textwidth]{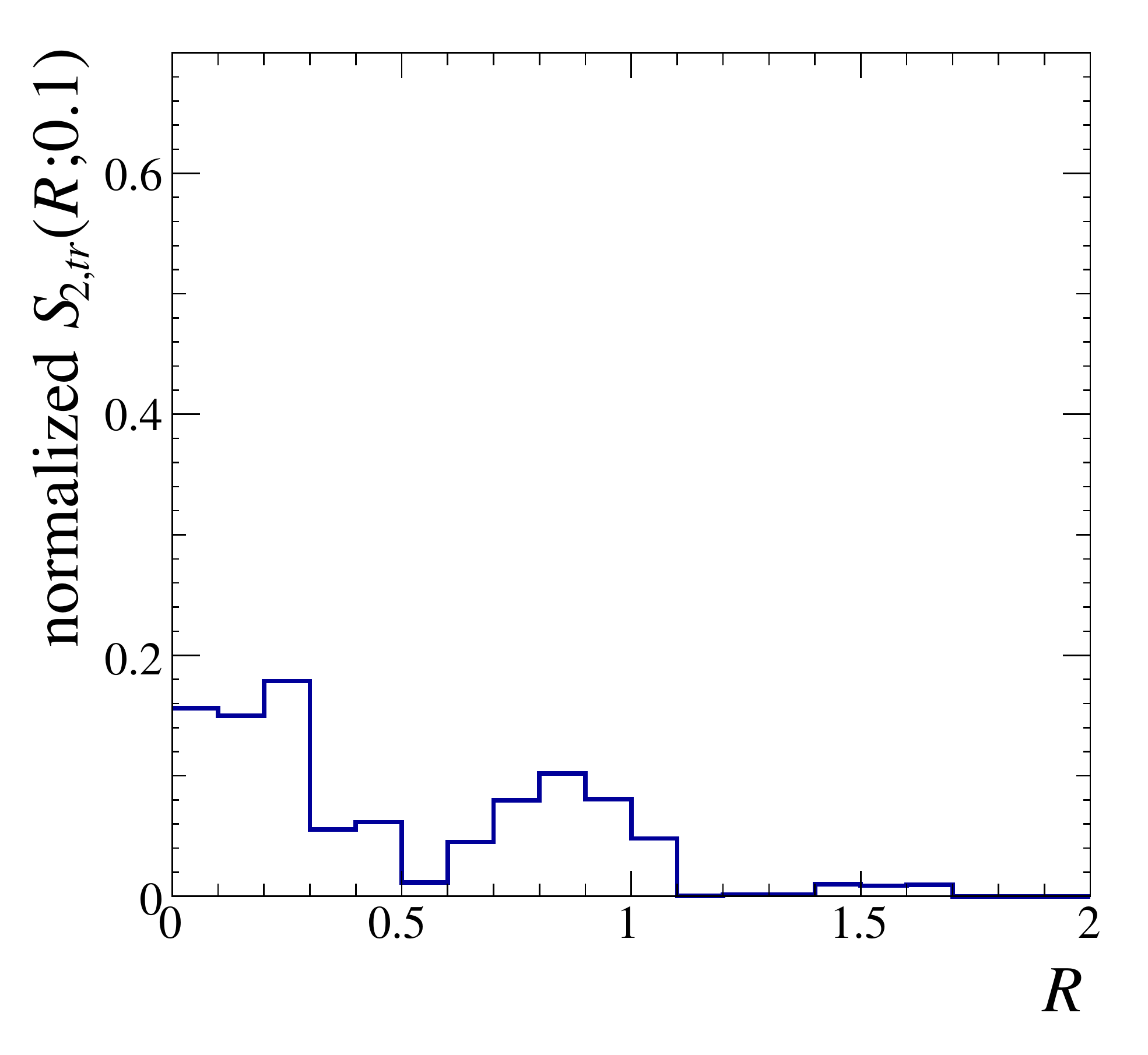}	
\end{center}
\caption{
The Pixelated jet image (left), two-point spectrum $S_2(R;0.1)$ (center), and trimmed spectrum $S_2(R;0.1)$ (right) of a Higgs jet which is Higgs-like only in the $\ANN{S_2+tr}$.
The sum of the bins of each spectrum is normalized to 1. 
The blue triangles and green plus symbols in the pixelated jet images indicate direction of $b$ quarks and gluons obtained from the matrix element including order $\alpha_S$ radiations.
We show the Higgs-like probability $p_h(Y_X)$ in an ANN model $\ANN{X}$ defined by \eqref{eqn:prob_higgs-like}.
}
\label{fig:jet_profile_char_onlyS2tr}
\end{figure}

The third example in \figref{fig:jet_profile_char_onlyS2tr} is the case where only $\ANN{S_2+tr}$, which takes into account trimmed $S_2(R)$, classifies the jet as a Higgs jet. 
This jet has a two-prong substructure but deeply buried in radiations compared to \figref{fig:jet_profile_HighD2_LowS2}.
As a result, $D_2$ is large, and $S_2(R)$ spectrum is falling toward high angular scale $R$ as in \figref{fig:jet_profile_char}.
Trimming helps $\ANN{S_2+tr}$ this time because $\ANN{S_2+tr}$ recognizes hard and soft substructure separately by comparing $S_2(R)$ and $S_{2,tr}(R)$.
Trimming does not change the tail of $S_2(R)$ distribution, which means the substructure at large $R$ is hard.

\section{Discussion and Conclusion}
\label{sec:5}

In this paper, we have introduced a spectral analysis of jet substructure with the artificial neural networks (ANN).
Unlike the other ANN approach, our algorithm use the spectral function $S_2(R)$ constructed from $p_T$ and $R$ of the pair of particles in the jet.
The spectrum $S_2(R)$ is useful in describing substructures with large angular separation by relatively small inputs.
ANN can learn non-local correlations in jets from the spectrum.
To show this, we have constructed ANN from $S_2(R)$, $\ANN{S_2}$, and compare it with ANN from $D_2$, $\ANN{D_2}$.
We have shown that $\ANN{S_2}$ discriminates between boosted Higgs jets and QCD jets with better performance compared to $\ANN{D_2}$.
Introducing trimming to $S_2(R)$ further helps ANN separate hard and soft substructures, and the ANN with trimmed observable outperforms the ANN without trimming.
The improvement comes from the better handling of the cases with radiation from $b$ parton or with contamination of other hadronic activities.

The improvement we observe is not large, because $D_2$ catches the two-prong substructure of the Higgs jet efficiently, but ANN analysis with $S_2(R)$ has much wider application.
One of the merits of $\ANN{S_2}$ and $\ANN{S_2+tr}$ is that the analyses automatically take care of radiations from the $b$ jet.
Note that the existence of radiation has to be taken care of even in the original mass drop tagger by \cite{Butterworth:2008iy}.
The $S_2(R)$ has information on three-point correlation and higher simultaneously and additional selections are not required.
Consequently, $S_2(R)$ can be used in a cascade decay of a heavy particle, especially the top quark.
We also note that $S_2(R)$ is sensitive to the color of the boosted heavy particle.
We will show in a separate publication that a modified $\ANN{S_2}$ discriminate color octet resonance and color singlet resonance efficiently \cite{octet}.

\begin{acknowledgments}
The authors would like to thank Amit Chakraborty, Patrick T. Komiske, Devdatta Majumder, and Tilman Plehn for useful discussions.
This work was supported by the Grant-in-Aid for Scientific Research on Scientific Research B 
(No.16H03991, 17H02878 [MMN]) and Innovative Areas (16H06492 [MMN]), and by World Premier 
International Research Center Initiative (WPI Initiative), MEXT, Japan. 
The work of SHL was supported in part by MEXT KAKENHI Grant Number JP16K21730.
\end{acknowledgments}

\bibliographystyle{JHEP}
\bibliography{JetSubstructureSpectroscopy}

\end{document}